\documentclass[usenatbib]{mn2e}
\input{psfig}

\newcommand{\beq}{\begin{eqnarray}}
\newcommand{\eeq}{\end{eqnarray}}
\newcommand{\gal}{{\sc galics}}
\newcommand{\mo}{{\sc momaf}}

\newcommand{\dd}{{\textrm d}}
\newcommand{\hatton}{{\sc galics~i}}
\newcommand{\devriendta}{{\sc galics~ii}}
\newcommand{\blaizota}{{\sc galics~iii}}
\newcommand{\blaizotb}{{\sc galics~v}}
\newcommand{\devriendtb}{{\sc galics~iv}}
\newdimen\hssize
\newdimen\htsize
\newdimen\hdsize
\newcommand{\apj}{ApJ}

\newcommand{\aj}{AJ}
\newcommand{\mnras}{MNRAS}
\newcommand{\aap}{A\&A}
\newcommand{\aaps}{A\&AS}

\newcommand{\pasp}{PASP}
\newcommand{\physrep}{Phys.~Rep.}

\begin{document}

\title[MoMaF]
      {MoMaF : the Mock Map Facility}
\author[Blaizot et al.]
      {J\'er\'emy Blaizot$^{1,2,3}$, Yogesh Wadadekar$^{1,4}$, Bruno Guiderdoni$^1$, St\'ephane T.
       Colombi$^1$, \and Emmanuel Bertin$^1$, Fran\c{c}ois R. Bouchet$^1$, Julien E.G. Devriendt$^{2,5}$
\& Steve Hatton$^1$\\
$^1$ Institut d'Astrophysique de Paris, 98 bis boulevard Arago, 75014 Paris, France. \\
$^2$ Oxford University, NAPL, Keble Road, Oxford OX1 3RH, United Kingdom.\\
$^3$ LAM, Traverse  du Siphon-Les trois Lucs BP8-13376 Marseille Cedex 12, France.\\
$^4$ Space Telescope Science Institute, 3700 San Martin Drive, Baltimore MD 21210, USA.\\
$^5$ CRAL, Observatoire de Lyon, 69561 St Genis Laval cedex, France\\}
\date{}
\pagerange{\pageref{firstpage}--\pageref{lastpage}}
\pubyear{2004}
\maketitle
\label{firstpage}
\begin{abstract}
We present the {\it Mock Map Facility}, a powerful tool to convert
theoretical outputs of hierarchical galaxy formation models into
catalogues of virtual observations. The general principle is
straightforward : mock observing cones can be generated using
semi-analytically post-processed snapshots of cosmological N-body
simulations. These cones can then be projected to synthesise mock sky
images. To this end, the paper describes in detail an efficient
technique to create such mock cones and images from the \gal{}
semi-analytic model, providing the reader with an accurate
quantification of the artifacts it introduces at every step. We show
that replication effects introduce a negative bias on the clustering
signal -- typically peaking at less than 10 percent around the
correlation length. We also thoroughly discuss how the clustering
signal is affected by finite volume effects, and show that it vanishes
at scales larger than about a tenth of the simulation box size. For the
purpose of analysing our method, we show that number counts and
redshift distributions obtained with \gal{}/\mo{} compare well to
$K$-band observations and to the 2dFGRS. Given finite volume effects, we also show that
the model can reproduce the APM angular correlation function. The
\mo{} results discussed here are made publicly available to the
astronomical community through a public database. Moreover, a
user-friendly Web interface ({\tt http://galics.iap.fr}) allows any
user to recover her/his own favourite galaxy samples through simple
SQL queries. The flexibility of this tool should permit a variety of
uses ranging from extensive comparisons between real observations and
those predicted by hierarchical models of galaxy formation, to the
preparation of observing strategies for deep surveys and tests of data
processing pipelines.

\end{abstract}
\begin{keywords}
astronomical data bases:miscellaneous - galaxies:statistics - large-scale structure of Universe - methods:numerical
\end{keywords}

\section{Introduction}
Large galaxy surveys in which homogeneous datasets
are acquired and analysed stand to benefit from the availability of mock images of
galaxy wide/deep fields. These images are very useful to design
observational strategies, and to put predictions of various models
into a format that can be directly compared to actual observations.
Such an activity of ``sky simulation'' is now running extensively as
part of the data processing centres for satellite missions and for the
next generation of ground-based instruments, because it is now
acknowledged that data processing and mock observations have to be
integrated into instrument building from the beginning of the project
to the interpretation of the data. However, such  sky simulations are
not easy, because they have to meet a certain level of realism to be
useful.  The purpose of this paper is to demonstrate how {\it
realistic} galaxy catalogues and mock images can be synthesised from
the outputs of a model of hierarchical galaxy formation.

Of course, mock galaxy surveys can be generated simply by drawing
galaxy types, luminosities and sizes from their distribution
functions. However, such an approach does not meet the requirements,
because (i) evolution cannot be included except in a very crude way,
especially if we are to mimic number evolution, (ii) multi-wavelength
surveys cannot be addressed easily, and (iii) spatial information
cannot be addressed since galaxy positions are only known with a
Poissonian distribution. Another approach is to use observed fields
rescaled on-purpose \citep{BouwensBroadhurstSilk98}. A third approach
is to start from theoretical priors and use numerical simulations to
describe hierarchical clustering. In this approach, the main issue is how
to transform mass to light and implement a robust method to get
galaxies within dark matter structures.

There are basically two paths that this theoretical approach can take
: (i) biasing schemes, and (ii) halo models. The simplest approach (i)
consists of ``painting'' galaxies on dark matter (DM) simulations
using phenomenological prescriptions to pick DM particles to be
galaxies. Such methods, based on the linear bias formalism use the
smoothed density field only \citep{ColeEtal98} and are thus very
efficient for big, low-resolution simulations \citep[e.g. the Hubble
Volume simulations described by][]{EvrardEtal02}. The more subtle approach (ii) uses halos
identified in DM simulations. Several variations exist. In the
simplest implementation, halos are populated with galaxies according
to a given halo occupation distribution
\citep{PeacockSmith00,ScoccimarroEtal01,BerlindWeinberg02} which may
depend on luminosity \citep{YangEtal04}. In a more sophisticated
approach, a Monte Carlo scheme is used to build a merger history tree
for each halo identified in the simulation, and a semi-analytic model
(SAM) is used to evolve galaxies in these trees
\citep[e.g.][]{KauffmannNusserSteinmetz97,BensonEtal00}. In this
approach, the SAM is used to generate a physically motivated halo
occupation distribution. Eventually, a hybrid approach can be used,
which extracts halo merging history trees from the dark matter
simulations, and use a SAM to evolve galaxies in these
\citep[][hereafter
\hatton{}]{KauffmannEtal99a,HellyEtal03a,HattonEtal03}. Mock
catalogues made using this technique have been made to mimic the CfA
redshift survey \citep{DiaferioEtal99} or the DEEP2 survey
\citep{CoilDavisSzapudi01}. As a matter of fact, all the
implementations of the two above methods are closely linked because
the bias formalism often relies on analysis of the SAMs themselves
\citep[e.g.][]{ColeEtal98,SomervilleEtal01}. They must therefore be
considered as complementary rather than competing. The spirit of the
physically motivated semi-analytic recipes can be extended to other
objects such as X-ray clusters \citep{EvrardEtal02}. However, these
approaches have generally been designed to fulfil the needs of a
specific survey (for instance 2dF for \citet{ColeEtal98}, or DEEP2 for
\citet{CoilDavisSzapudi01}). This limited scope can be extended in at
least three ways.

First, it would be interesting to elaborate a generic approach to
address the construction of mock observing cones from the outputs of
N-body simulations at various cosmic times. The main issue is that,
depending on the depth and solid angle of the observing cone, the
finite size of the box may call for box replication along the
line-of-sight (hereafter {\it radial replication}), and box
replication perpendicularly to the line of sight, at the same cosmic
time (hereafter {\it transverse replication}). Wide field, shallow
surveys, with negligible evolution, can be constructed mainly from a
single box (the last output of a simulation corresponding to
$t=t_0$). In contrast, deep, pencil-beam surveys generally have to use
numerous radial replications, whereas they may avoid transverse
replication. Several issues have to be addressed here : the effects
replication might have on mock catalogues, the effect using a finite
volume might have on catalogues, and the sensitivity of catalogues to
the number of time outputs of the root simulation. Of course, using a
larger box size would improve the situation, but given finite computer
resources (CPU time and memory), using a larger box would require a
trade-off in the mass resolution of the simulation, which is not
acceptable if galaxies are to be modelled with a sufficient level of
realism. A Hubble volume would be the ideal situation avoiding any
radial or transverse replication, but so far, the largest volume
simulation ($\Lambda$CDM with $3000 h^{-1}$ Mpc on a side, and $10^9$
particles) has a particle mass of $m_p=2.25 \times 10^{12}h^{-1}
M_\odot$ \citep{EvrardEtal02}, much too large to address galaxy
formation with any of the ``hybrid models''. For instance, the
$\Lambda$CDM simulation used in \hatton{} has only $100 h^{-1}$ Mpc on
a side, and $m_p=5.51 \times 10^{9}h^{-1} M_\odot$; yet resolution
effects are visible for galaxies fainter than $L^*/8$ at $z=0$. While
we await a three orders of magnitude improvement of the simulations,
addressing the replication issues is unavoidable if one wants mimic
large-volume observations with high resolution.

Second, the mock catalogues are useful if they gather together a large
number of potentially observable properties. For instance, it is
obvious that a mock catalogue designed to prepare and analyse a
redshift survey of a magnitude-limited sample in a given photometric
band, will incorporate at least the predicted redshifts and apparent
magnitudes in that band. But the redshift survey will also be used for
follow-up at other wavelengths, and other studies (for instance
spectral classification once the spectra are properly calibrated). A
good mock catalogue will be able to provide all these pieces of
information at wavelength bands different from those of the original
survey. Ultimately the mock catalogue will enable the production of
field images at many wavelengths, making source extraction using the
same data processing pipeline as the actual observations possible.

Third, the catalogues quickly become very large, and the question of
accessibility to relevant information becomes crucial. Generally they
are made available on Web pages as ASCII tables, mostly as galaxy
catalogues from snapshots, more rarely as galaxy catalogues from
observing cones. The more realistic these tables try to be (by
including many galaxies with many properties), the more difficult to
read and use they become, because of their growing size. The solution
to this problem is to make the catalogues accessible through a database
that can be queried to make ad hoc sub-samples fitting specific needs
within a wide range of possibilities.

The purpose of this paper is to contribute along these three lines, by
(i) presenting a package called \mo{} (for {\it Mock Map Facility})
that generates observing cones from the outputs of our \gal~model, and
(ii) discussing in detail the limitations of the method. From these
observing cones, synthetic catalogues are generated, that can be
easily related to the catalogues of galaxies in the snapshots. The
catalogues gather together a large number of properties, including
magnitudes in many photometric bands of interest. These
\gal/\mo~catalogues are made available in an on-line database that can
be queried through a simple Web interface at {\tt
http://galics.iap.fr}.

To illustrate our technique, we use examples drawn from the
$\Lambda$CDM simulation and the \gal~ post-processing described in
\hatton. This simulation is a compromise in terms of mass resolution
and volume size, and gives a satisfactory description of the
luminosity functions over typically 5 magnitudes. However, the
techniques we describe are generic, and can be used for larger
simulations. In this study, we do not address in detail the drawbacks
of our model (see \hatton) in terms of mass resolution or limited
volume, nor the quality of its predictions. We are only interested in
how these predictions can be converted into mock observations. First
examples of using these predictions and the mock catalogues can be
found in Devriendt et al., 2004 (hereafter \devriendta, in
preparation), \citet[][hereafter \blaizota]{BlaizotEtal04}. Two other
papers will address issues which are more relevant to mock images:
multi-wavelength faint galaxy counts (\devriendtb), and correlation
functions (\blaizotb).

This paper is organised as follows. Section 2 summarises the main
features of the \gal{} model that are relevant to our study. In
section 3, we describe our technique of catalogue and map building
from the simulation snapshots. In Section 4, we explore the different
limitations of our method, most of which are actually general enough
to apply to other mock catalogues in the literature based on the
tiling method. We explain in section 5 how all the products of \gal{}
and \mo{} are stored in a relational database accessible from the web,
and illustrate a few key features of this database. Section 6 contains
a discussion about how these mock catalogues and images may be used,
and presents perspectives for further developments.

\section{The GALICS model}

\gal~(for Galaxies In Cosmological Simulations) is a model of
hierarchical galaxy formation which combines high resolution
cosmological simulations to describe the dark matter content of the
Universe with semi-analytic prescriptions to deal with the baryonic
matter. This hybrid approach is fully described in \hatton{} and
\devriendta{} and we only briefly recall its relevant features here.

\subsection{Dark matter simulation}
The cosmological N-body simulation we refer to throughout this paper
was done using the parallel tree-code developed by \citet{Ninin99}. It
is a flat cold dark matter model with a cosmological constant
($\Omega_m = 0.333$, $\Omega_{\Lambda} = 0.667$). The simulated volume
is a cube of side $L_{box}=100h_{100}^{-1}$Mpc, with $h_{100}= 0.667$,
containing $256^3$ particles of mass $8.272\times 10^9$M$_{\odot}$,
with a smoothing length of 29.29 kpc. The power spectrum was set in
agreement with the present day abundance of rich clusters \citep[$\sigma_8 =
0.88$, from][]{EkeColeFrenk96}, and we followed the DM density field from z=35.59 to z=0,
outputting 100 snapshots spaced logarithmically in the expansion
factor.

In each snapshot we use a friend-of-friend algorithm to identify
virialised groups of more than 20 particles, thus setting the minimum
dark matter halo mass to $1.65\times 10^{11}$ M$_{\odot}$. We compute a set
of properties of these halos, including position and velocity of the
centre of mass, kinetic and potential energies, and spin
parameter. Then, assuming a density profile for the virialised dark
matter, we compute the virial radius a spherical halo would have to
have the same mass and potential energy, thus making the link to the
idealised semi-analytic approach.

Once all the halos are identified in each snapshot, we compute their
merging history trees, following the constituent particles from one
output to the next one. The merging histories we obtain are by far
more complex than in semi-analytic approaches as it includes
evaporation of halos, fragmentation, and several artifacts due to
loose friend-of-friend identifications. The way we deal with these is
described in detail in \hatton.

\subsection{Baryonic Prescriptions, or how mass turns into light} \label{sec:baryonic prescriptions}
When a halo is first identified, it is assigned a mass of hot gas,
assuming a universal baryonic to dark matter mass ratio ($\Omega_b =
0.045$ in our fiducial model). This hot gas is assumed to be shock
heated to the virial temperature of the halo, and in hydrostatic
equilibrium within the dark matter potential well. The comparison of
the cooling time of this gas to its free-fall time, as a function of
the radius, yields the mass of gas that can cool to a central disc
during a time-step. The size of this exponential disc is given by
conservation of specific angular momentum during the gas in-fall and
scales as the spin parameter of the halo. Then, the cooled gas is
transformed into stars with a rate proportional to its mass divided by
the disc dynamical time, with a given efficiency. The stars formed are
distributed in mass according to an initial mass function (IMF) taken
from \citet{Kennicutt83}. The stellar population of each galaxy is
then evolved between the time-steps, using a sub-stepping of at most 1
Myr. During each sub-step, stars release gas and metals in the ISM,
and we follow this gas recycling in time, assuming instantaneous
mixing. The massive end of the stellar population shortly explodes
into supernovae which also release metals and energy in the ISM or in
the IGM. We model this as a function of the instantaneous star
formation rate.

When two halos merge, the galaxies they contain are gathered within
the same final halo and their orbits perturbed. Subsequently, due to
dynamical friction or satellite-satellite collisions, they can
possibly merge. A ``new'' galaxy is then formed (the {\it descendant}
of the two {\it progenitors}) and the stars and gas of the progenitors
are distributed in three components~: a disc, a bulge, and a
starburst, the amount of what goes where being fixed by the ratio of
masses of the two progenitors. The new galaxy can be elliptical (in
shape) if the two progenitor galaxies have about the same mass, or
remain a spiral if one of the merged galaxies has negligible mass.

The spectral energy distributions (SEDs) of our modelled galaxies are
computed by summing the contribution of all the stars they contain,
according to their age and metalicity, both of which we keep track
of all along the simulation. Then, extinction is computed assuming a
random inclination for disc components, and the emission of dust is
added to the extinguished stellar spectra with {\sc stardust}
\citep{DevriendtGuiderdoniSadat99}. Finally, a mean correction for
absorption through the intergalactic medium (IGM) is implemented
following \citet{Madau95}, before we convolve the SEDs with the
desired filters in the observer frame.

\subsection{Resolution effects} \label{sec:resolution}
The mass resolution of the DM simulations affect both the physical and
statistical properties of modelled galaxies in the following ways :
\begin{enumerate}
\item The particle mass of the cosmological simulation sets a minimum
halo mass. Converting this halo mass into a galaxy mass, assuming that
all the gas in the halo cools, one gets a threshold mass above which
our sample of galaxies is complete : the {\it formal completeness
limit}. Below this mass, although we do have galaxies, our sample is
not complete since we miss galaxies in undetected halos. This direct
effect of mass resolution is responsible for the lack of dwarf
galaxies in the standard \gal{} model. To express the completeness
limit in terms of magnitudes is not straight-forward because of the
complex processes that convert mass into light. One can define a
limiting magnitude, at a given redshift, such that, say, $95\%$ of the
galaxies brighter than that will be more massive than the formal mass
resolution. Because there is no one-to-one relation between mass and
luminosity, however, the luminosity selection is in practice more
drastic than the selection on mass. As an example, these magnitudes
are given in several wave-bands at $z=0$ and $z=3$, in Table
\ref{table:limit mags}. They can easily be derived for other wave-bands
or redshifts from the \gal{} database (see section
\ref{sec:database}).

\item In a Universe dominated by {\it cold} dark matter, small structures form first, and then merge and accrete material so as to evolve into larger haloes. In other words, the characteristic mass $M_*$ of the mass distribution of haloes increases as redshift decreases. The mass resolution of our numerical simulation is fixed, however, and does not allow us to identify objects less massive than $\sim 1.6\times 10^{11}M_\odot$, at any redshift. Hence, going back in time, more and more haloes are not resolved, and one eventually reaches a point where no halo can be detected. We call $z_{lim}$ the limit redshift when this happens. At higher redshifts, we miss all possible galaxies. In our simulation, one find $z_{lim}\sim 7$.

\item A more subtle effect of resolution is that missing small
structures means missing part of galaxies' histories. In practice, we
showed that for our standard simulation, a galaxy needs to have
evolved for about 1 Gyr before its properties have converged
(see \blaizota). Although this is virtually no constraint at $z=0$, where
most galaxies are much older than 1 Gyr, the constraint becomes
drastic at $z=3$, when the age of the universe is only about 2 Gyr.
To ease the selection of mature galaxies for users of the database
(see section \ref{sec:database}), we assign the morphological type
'Im' to immature galaxies.
\end{enumerate}

\begin{table}
\begin{center}
\begin{tabular}{lccc}
\hline \hline
        & Wave-band & 95 \% completeness & 75 \% completeness \\
\hline 
        & U        & -19.6 mag          &  -18.2 mag  \\
        & B        & -19.5 mag          &  -18.5 mag  \\
        & V        & -20.1 mag          &  -19.4 mag  \\
$z = 0$ & R        & -20.6 mag          &  -20.1 mag  \\
        & I        & -21.1 mag          &  -20.7 mag  \\ 
        & J        & -21.8 mag          &  -21.5 mag  \\
        & K        & -22.8 mag          &  -22.5 mag  \\
\hline		     			  			          
        & $U_n$    & -20.6 mag          &  -20.3 mag  \\
$z=3$   & $G$      & -21.1 mag          &  -20.9 mag  \\
        & $R$      & -21.1 mag          &  -20.9 mag  \\
\hline
\end{tabular}
\end{center}
\caption{95\% and 75\% completeness limits in terms of absolute
rest-frame magnitudes at redshifts 0 and 3. At $z=0$, the magnitudes
are expressed in the Vega system, and the filters are Johnson's. At
$z=3$, the magnitudes are expressed in the AB system, and the filters
are those from \citet{SteidelHamilton93}.}
\label{table:limit mags}
\end{table}


\section{Mock Observations}
In this section, we explain how we convert the outputs of \gal{}
described above into mock observations. We first describe the inputs
we need from \gal{} or any other model/simulation of galaxy
formation. Then, we show how these inputs are turned into mock maps,
and point out the main limitation of our technique : replication
effects. Finally, we briefly explain how we can project catalogues
onto realistic pre- or post-observing maps.

\subsection{Inputs}
The method we developed to generate mock catalogues from outputs 
(or {\it snapshots}) of cosmological simulations at a finite number 
of redshifts is general and can be used for a variety of objects 
(e.g. clusters or quasars). Here, we describe the features needed in
these snapshots for galaxies.

The snapshots have to be (cubic) volumes of equal comoving size (in
our standard simulation, $L_{box} = 100$ $h^{-1}$Mpc) with periodic
boundary conditions. These snapshots must each contain the following
information :
\begin{itemize}
  \item The redshift, or expansion factor of the snapshot.
  \item The position of each galaxy within the snapshot.
  \item The velocity of each galaxy within the snapshot.
  \item The characteristic scale-length of each component of each galaxy (disc, bulge and
    burst).
  \item The inclination of each galaxy, which was used to compute its
    extinction.
  \item The absolute AB magnitude of each galaxy in the desired filters,
    computed in the observer frame as 
    \beq \label{eq:magabs}
    M_{\nu_0}(z) = -2.5 \log\left( \frac{L_{\nu_0(1+z)}}{[10 \textrm{
    pc}]^2} \right) -2.5 \log(1+z) + 48.6, 
    \eeq 
    where $z$ is the redshift of the snapshot, and 
    \beq 
    L_{\nu_0(1+z)} = \int (1+z)f_{\nu_0}(\nu)L[\nu(1+z)]\textrm{d}\nu 
    \eeq 
    is the luminosity of a galaxy at redshift $z$, through a
    normalised filter response $f_{\nu_0}$.  Note that because the
    peculiar velocities or positions relative to the observer are not
    known at this stage, we use the redshift $z$ of the snapshot to
    compute these magnitudes. This approximation will be corrected for
    when we compute apparent magnitudes (see
    Sec. \ref{sec:app_props}). Also note that these magnitudes take
    into account extinction by the intergalactic medium, computed at
    the redshift of the snapshot.
  \item The first order derivatives of the above magnitudes with
    redshift, in each filter. For galaxies in snapshot $i$, these
    derivatives are estimated as
    \beq \label{eq:dmdz}
    \frac{\dd M}{\dd z} = \frac{M[z(i-1)] - M[z(i)]}{z(i-1)-z(i)}, 
    \eeq
    where $z(i)$ is the redshift of snapshot $i$ (in our convention,
    $z(i)<z(i-1)$), and $M[z]$ the observer-frame absolute magnitude
    assuming the galaxy is at redshift $z$ (Eq. \ref{eq:magabs}). Note
    that this expression does not account for the evolution of
    galaxies, as it involves the magnitudes of the {\it same} galaxy
    put at different redshifts. Eq. \ref{eq:dmdz} however captures
    K-correction and variations of IGM extinction with $z$, which are
    the main drivers of average variations of apparent properties with
    redshift in mock catalogues (see Sec. \ref{sec:app_props} and
    \ref{sec:fte}).

\end{itemize}

All the above quantities are direct outputs of \gal{}, except for
positions and velocities. These are computed as a post-treatment,
using information from the DM simulations (positions and velocities of
the halos) and from \gal{} (orbital radii). The position of a galaxy
within a snapshot is thus defined by $\vec{g} = \vec{h} +
r_{orb}\times \vec{u}$, where $\vec{h}$ is the position of its host
halo, $r_{orb}$ the orbital radius of the galaxy, and $\vec{u}$ a
normalised vector of random direction. The peculiar velocity of a
galaxy is defined as $\vec{v}_g = \vec{v}_h + \delta\vec{v}$, where
$\vec{v}_h$ is the peculiar velocity of its host halo, and
$\delta\vec{v}$ is the peculiar velocity of the galaxy within this
halo. The amplitude of $\delta\vec{v}$ is drawn randomly from a
Gaussian distribution of width equal to the circular velocity of
the halo, and its direction is random. Note that the velocities of
central galaxies are taken to be that of the centre of mass of their
host haloes.

\subsection{Mock Catalogues}
\begin{figure*}
\begin{tabular}{cc}
\psfig{figure=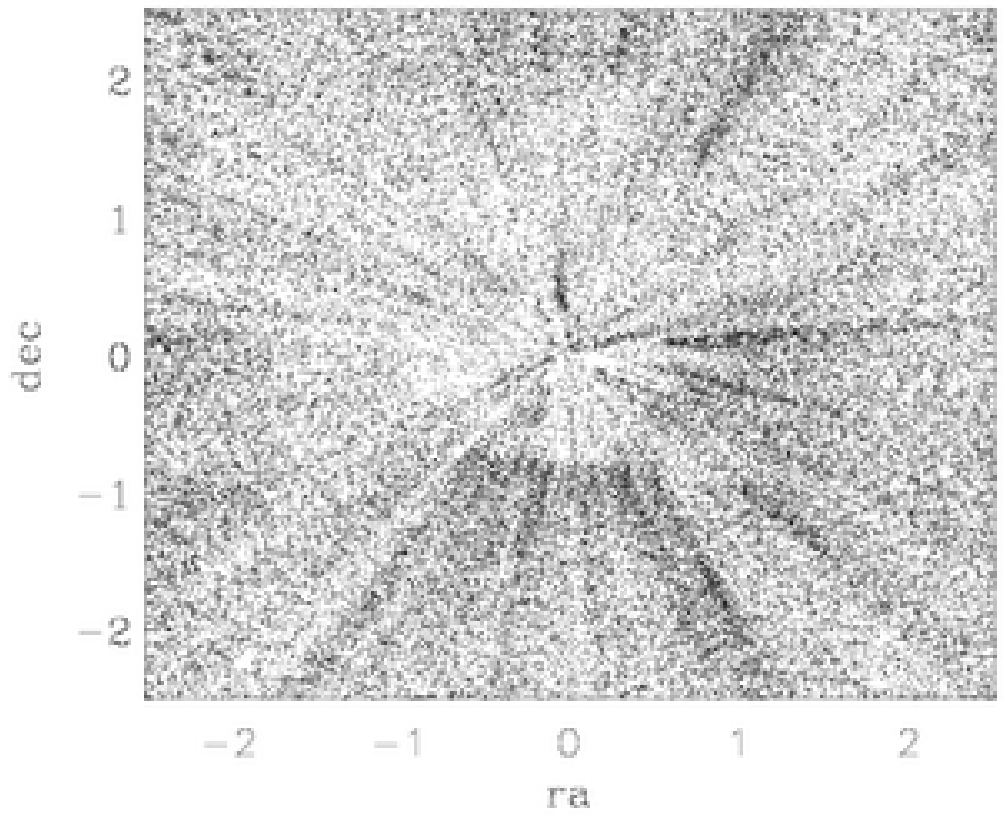,width=\hssize}&
\psfig{figure=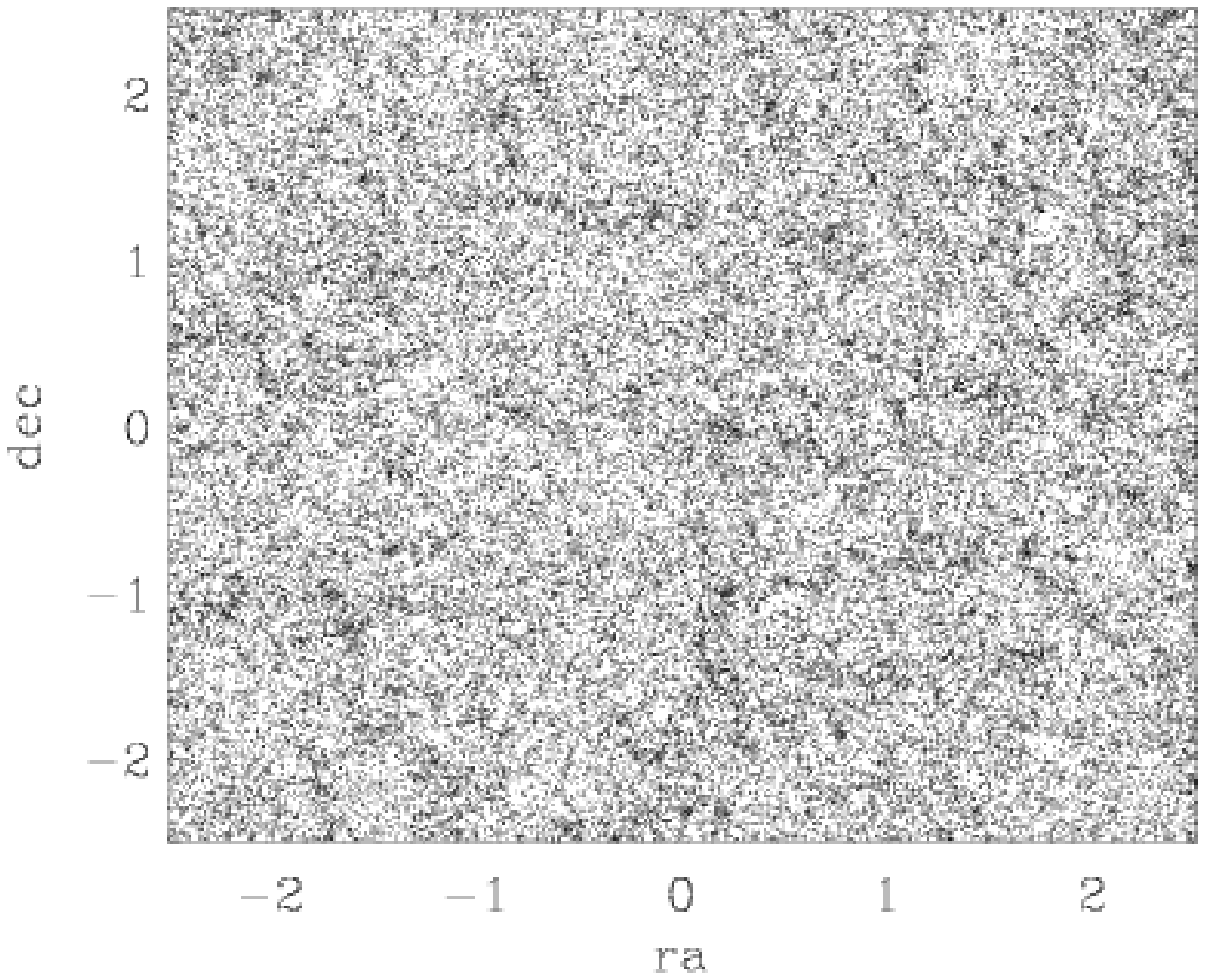,width=\hssize}
\end{tabular}
\caption{Angular projection of mock catalogues of $5\times5$ square
degrees (the ``r.a.'' \& ``dec.'' labels are arbitrary labels for two
orthogonal directions on the mock sky). Each point represents a bright
galaxy. The left hand side panel shows a catalogue in which the
snapshots were piled together without any randomisation. This leads to
replication effects, similar to a perspective effect (in an
expanding/evolving universe). The right hand side panel shows a
catalogue containing the same galaxies, which was made using the
random tiling method. All replication effects have disappeared.}
\label{fig:star_wars}
\end{figure*}

Such inputs, corresponding to the same simulated region of universe at
different redshifts, will cause replication effects when piled in a
mock light-cone, namely the regular repetition of structures in mock
catalogues or images. {\it Transverse replications} are due to the
fact that the same volume, in the same state of evolution, is used
several times to fill an observing cone across the line of
sight. {\it Radial replications} occur because the same volume,
although taken at different cosmic times, is repeatedly used to fill
the observing cone along the line of sight. Because the largest
structures evolve slowly (i.e. over several time-steps), they will
create pseudo-periodicity in mock catalogues or mock maps. In
Fig. \ref{fig:star_wars} (see also Fig. \ref{fig:replication}), we
show how replications create an artificial perspective effect in
catalogues (left hand side panel). Replication effects can be suppressed
with the ``random tiling'' method (right hand side panel), which we
describe here.

\subsubsection{random tiling}

Building a catalogue from the inputs described above consists of
distributing the simulated galaxies in an observing cone, and
computing their apparent properties in this new geometry. First, we
define a three-dimensional pavement of cubic underlying boxes of side
$L_{box}$ ($=100$ comoving $h^{-1}$Mpc). Then, we fill the underlying
boxes inside the light cone with galaxies in the following way :
\begin{itemize}
\item determine the time-steps $i = n$, ..., $n + k$ which will be
  needed in order to fill the current underlying box, knowing that time-step
  $i$ will be used to fill the light cone between $[z(i-1)+z(i)]/2$
  and $[z(i)+z(i+1)]/2$;
\item to each of these snapshots, apply the same transformation, which is
   a random combination of the following transformations :
  \begin{itemize}
  \item a {\it shift} of random amplitude (between $0$ and
    $L_{box}$) in each of the three directions $(x,y,z)$,
  \item a {\it rotation} of $0$, $\pi/2$, $\pi$, or $3\pi/2$
    around each axis,
  \item the {\it inversion} of one of the axes picked randomly
    (e.g. $x \mapsto -x$), or none;
  \end{itemize}
\item use the transformed positions and velocities of galaxies to
  include them in the light cone and compute their apparent
  properties;
\item move on to the next underlying box, and repeat the previous steps
  until the light cone is filled.
\end{itemize}

\begin{figure}
\begin{tabular}{l}
\psfig{figure=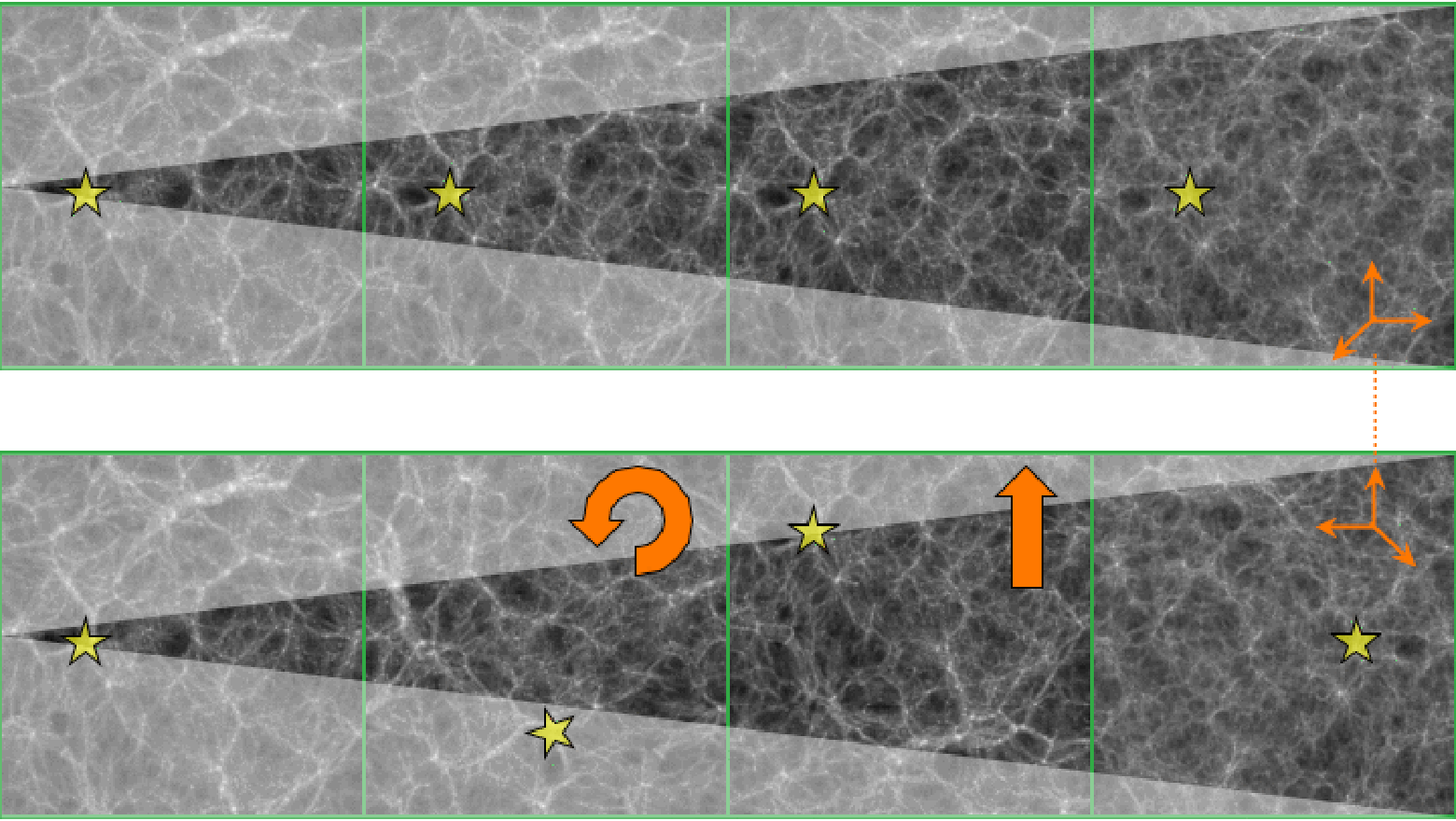,width=\hssize}\\
\psfig{figure=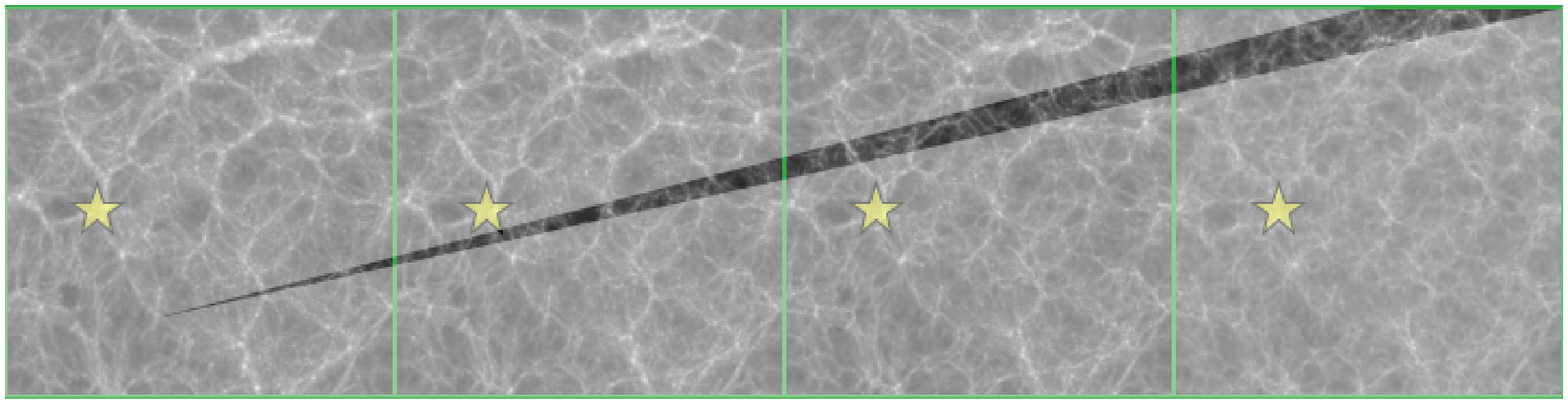,width=\hssize}
\end{tabular}
\caption{Illustration of the cone-making process. On the {\it top
panel} we show the result of a straight-forward tiling of
snapshots. In this case, the structures (such as the yellow star)
appear repeatedly along the line of sight. The {\it middle panel}
shows the effect of the three types of transformation we apply
randomly to each underlying box. Thanks to rotations, translations,
and inversions, the underlying boxes are decorrelated one from
another. On the {\it lower panel} we show how it is possible to avoid
re-shuffling underlying boxes to generate a pencil-beam type field.}
\label{fig:replication}
\end{figure}

The first step allows a galaxy in the cone to be taken from the output
box which has the closest redshift to the galaxy's redshift relative
to the observer. This has the advantage of picking galaxies at a stage
of evolution as close as possible to that they would have if we had
continuous outputs.

In the second step, the shifting, rotating and inverting of the
underlying boxes is done to suppress replication effects. The
shuffling of the underlying boxes, outlined with thick lines in figure
\ref{fig:replication}, decorrelates them from one another, thus
suppressing replication effects as well as any information on scales
larger than the box size.  Although breaking the continuity of the
density field makes us loose a fraction of spatial information (see
section \ref{sec:rtb}), we chose this solution because we have good
control on this information loss.

For deep pencil-beam surveys it is possible to avoid replication
effects simply by choosing an appropriate line of sight so that the
light-cone will intersect different regions of each underlying box. It
is better in this configuration not to shuffle the boxes so as to keep
all the spatial information. This is an option which is implemented in
our code, and illustrated on the lower panel of Figure
\ref{fig:replication}. Note however that, although the density field
is continuous throughout the cone, clustering information on scales
larger than $L_{box}$ is still missing, because it is not contained in
the DM simulations~: replication effects can be suppressed but not
finite volume effects (see Sec. \ref{sec:fve}).

It is also possible to chose the position of the observer, relative to
the first underlying box, in both options. This allows to test for
cosmic variance on local sources, and is useful to understand the
statistical significance of the bright end of galaxy counts.

\subsubsection{apparent magnitudes} \label{sec:app_props}
Because we use a finite number of time-steps, (i) galaxies are picked
at a cosmic time which is different from that corresponding to their
distance in the mock light cone, (ii) the SEDs are not convolved with
the filters at the exact redshifts, and (iii), IGM extinction is not
computed for the correct redshifts.  Point (i) means that an
individual galaxy is not taken at the stage of evolution it would have
in the case of continuous outputs. However, this does not affect the
statistical properties of the mock catalogues because the overall
galaxy population does not evolve much between time-steps, on
average. This issue is also discussed in Sec. \ref{sec:fte}.

Points (ii) and (iii), we correct for as follows.
We define corrected observer-frame absolute magnitudes as
\beq \label{eq:mag_correction} 
M_{cor} = M[z(i)] + \frac{\dd M}{\dd z} \times [z(d) - z(i)], 
\eeq 
where $ M[z(i)]$ is the observer-frame absolute magnitude computed by
\gal{} at redshift $z(i)$ of time-step $i$ (Eq. \ref{eq:magabs}),
$z(d)$ is the redshift of the galaxy evaluated from its comoving
distance $d$ to the observer in the mock light cone and taking into
account the peculiar velocity of the galaxy along the line of sight,
and $\dd M /\dd z$ is defined in Eq. \ref{eq:dmdz}.  Note that this
derivative only accounts for distance effects (K-correction and IGM
extinction) and not evolution (point (i) above). The apparent
magnitude of a galaxy is then obtained with the luminosity distance
$d_L$~:
\beq
m = M_{cor} + 5 \log \left( \frac{d_L}{10\textrm{ pc}} \right).
\eeq
Thanks to the first order correction of magnitudes, the distribution of
galaxies in apparent colour-colour plots is continuous. This is
especially important for colour selections of distant galaxies as shown in \citet{BlaizotEtal04}.

\begin{figure}
\psfig{figure=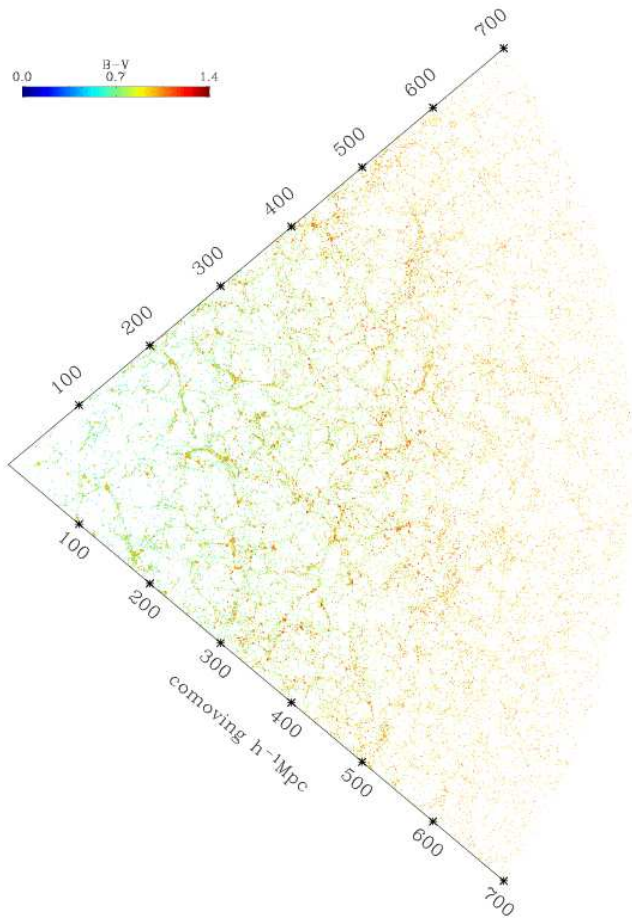,width=\hssize}
\caption{Mock 2dFGRS : each point represents a galaxy brighter than
$b=19.5$ from a light-cone of $75\times 4$ square degrees. The cone
was truncated at a comoving distance of $700h^{-1}$Mpc, corresponding
to $z\sim 0.23$. The colours of the points indicate the apparent $B-V$
colours of the galaxies according to the colour table in the upper-left
corner.}
\label{fig:ex_light_cone}
\end{figure}

In Fig. \ref{fig:ex_light_cone}, we show an example light cone with a
detection limit close to that of the 2dFGRS \citep{CollessEtal01}. Each point
represents a galaxy with $b < 19.5$, and the colours indicate apparent
$B-V$ colour of the galaxies.

\subsection{Mock Maps}
Two types of maps are useful to address different issues~:
\begin{itemize}
\item {\it pre-observation maps} are a simple projection of a mock
catalogue on the sky. The only additional assumption required here is
the functional form for the galaxy light profiles (e.g. an exponential
disc).
\item {\it post-observation maps} include, in addition, realistic
modelling of the characteristics of the telescope/detector combination
(e.g. diffraction effects, readout noise, photon shot noise). Where
appropriate, atmospheric effects can also be included (e.g. seeing,
air glow). SkyMaker \citep{ErbenEtal01} is a useful tool for producing post observation maps.
\end{itemize}

\subsubsection{Pre-observation maps}\label{sec:pre-obs}
Consistent with the modelling of galaxies in \gal, we display
disc components with an exponential profile, and bulges and starbursts
with a Hernquist profile \citep[][equations 32-34]{Hernquist90}. The
profiles are truncated at about ten times the component's half-mass
radius, and the Hernquist profile is also dimmed exponentially
starting at five times the half-mass radius. To gain speed, we build a
series of face-on disc and bulge templates on grids of different
resolutions (e.g. from $2\times 2$ to $2048 \times 2048$ pixels), each
one normalised to unity. When adding a galaxy's contribution to the
final map, we chose the template which has a resolution just above
that of the final map.  For bulges, as they are assumed to be
spherically symmetric, the projection is straightforward, and only
rescaling of the template to the component's size is required. For
discs, we first have to flatten the template to account for
inclination\footnote{This inclination is the same as that used in
\gal~to compute the extinction of light by dust.}, and then to rotate
it to map the disc's orientation. Eventually, we project the
transformed templates on the final map grid, multiplying each template
by the flux of the component it represents. The total flux on the
final map is thus the sum of the fluxes of all galaxies in the light
cone, except when they are truncated on the border of the image.

Some aliasing effects appear because of the projection of the tilted
template grids on the final map, but these will be washed out when the
map is convolved with a PSF afterwards. Since the aim of this tool is
to produce pre-observation maps (with a resolution that should be
higher than the final post-observation map), there is not much point
in correcting this effect via bilinear interpolation or other
CPU-expensive methods.

Note that the images produced this way are not limited in magnitude
(up to the resolution limit of the simulation) and include the
contributions of all galaxies in the cone. It is important that all
sources are added (even though some may be fainter than the detection
limit) for estimating the background intensity. This is particularly
relevant to far infrared or sub-millimetre surveys which are limited
by confusion and where the background contains a good part of the
information.

\begin{figure*}
\begin{tabular}{cc}
\psfig{figure=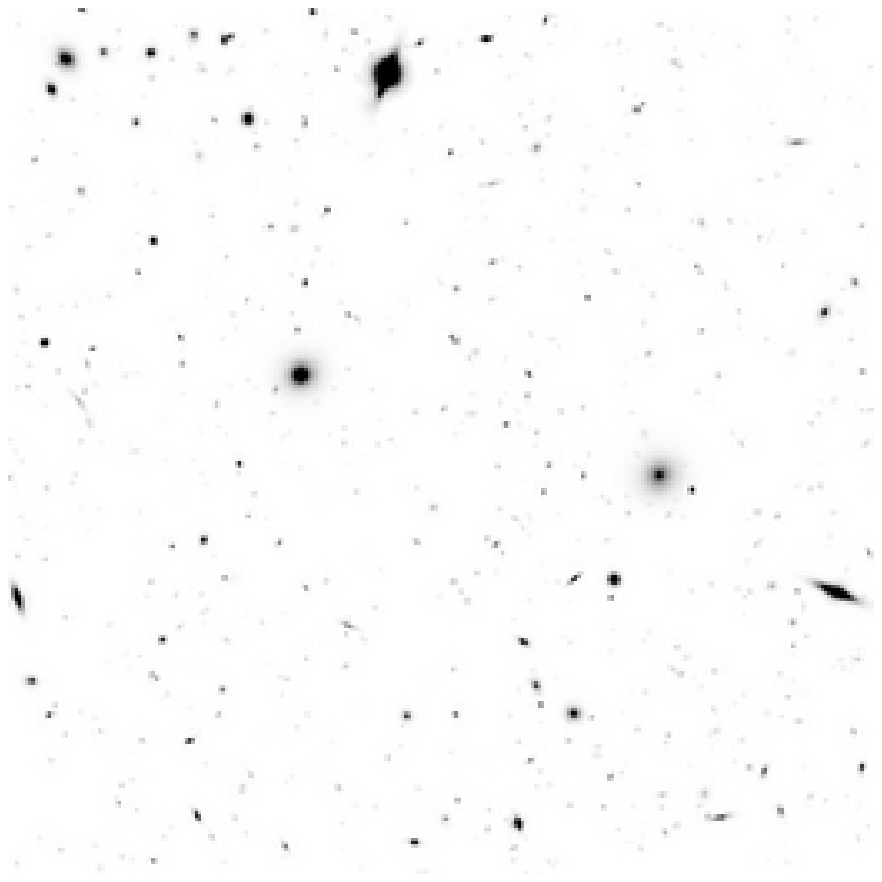,width=\hssize} & 
\psfig{figure=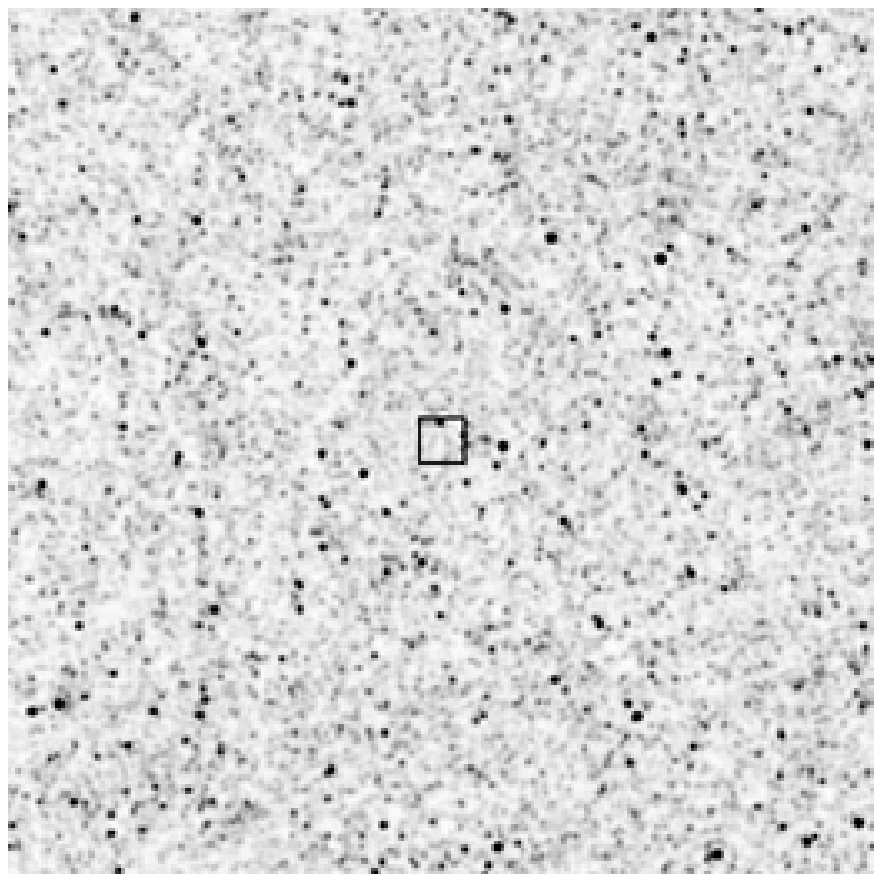,width=\hssize}
\end{tabular}
\caption{{\it Left hand side panel :} mock map of a $3\times 3$ arcmin$^2$ field, in the R band. {\it Right hand side panel :} mock map of a $1$ deg$^2$ field, at 170$\mu$m, convolved with a Gaussian PSF of width 10 arcsec. The central square in this image outlines the field shown in the optical in the left-hand side panel.}
\label{fig:mock_maps}
\end{figure*}

Example mock maps are shown in Fig. \ref{fig:mock_maps}. The left hand
side panel shows an optical view (R band) of a $3\times 3$ square
arcmin field, and the right hand side panel shows the far IR view of
the same field (at $170\mu$m). This latter image was convolved with a
Gaussian PSF of width 10 arcsec to mimic an observation by the
PACS instrument on-board Herschel. No noise was added to these mock
maps.

Finally, note that \mo{} allows to generate all-sky maps, using the HEALPix pixelisation \citep{GorskiEtal02} chosen by the Planck consortium.

\subsubsection{Post-observation maps}\label{sec:post_obs}

It is considerably more difficult to generate post observing maps
because separate modelling is required for each telescope/detector
combination. \mo{} is designed to feed Instrument Numerical Simulators
(INS) with realistic catalogues or pre-observation maps. In the
optical and near-infrared domain, a ready general tool for post
observing map generation is available in Skymaker
\citep{ErbenEtal01}. We briefly discuss this general INS here as an
example of \mo{} possibilities.

Skymaker is an image simulation program, originally designed to assess
SExtractor detection and measurement performances
\citep{BertinArnouts96}. The code (currently at version 2.3.4) has
been much improved since. It is capable of simulating star and galaxy
images with high level of accuracy. Galaxies in Skymaker are modelled
as a combination of a de Vaucouleurs bulge and an exponential
disk. Various sources of noise and convolution with the Point Spread
Function can be included as desired.
 
There are two input files required for Skymaker to generate an
image. One is a configuration file specifying the characteristics of
telescope and detector and the seeing conditions. The second is the
source list, which can include stars and galaxies.
A typical line for a galaxy in a source list for Skymaker includes the
``total'' magnitude, the bulge-to-total luminosity ratio, bulge
equivalent-radius in arc-second, projected bulge aspect ratio, bulge
position angle in degrees, disk scale length in arc-second, disk aspect
ratio and disk position angle in degrees. All of these are natural
outputs of \gal/\mo, as discussed in the previous section.
There is no provision for adding starburst components in Skymaker. We
use a workaround for such cases. We add the burst components from
\gal~as additional bulges, with a scale-length obtained from \gal,
a bulge-to-total luminosity ratio of 1.0 and the appropriate starburst
magnitude as the ``total'' magnitude.

\section{Limitations of the method}
Ideally, one should build a mock catalogue from a simulated volume
much larger than the light-cone, and output smoothly the physics by
propagating photons towards the observer through the expanding
simulated universe. Although such simulations are becoming feasible
today, their computational cost is still prohibitive. The method we
propose with \mo{} stems from the same philosophy as \gal{} and
consists of extracting as much information as possible from a given
simulation, and use that to build realistic catalogues at relatively
low computational expense. Of course, however sophisticated the
method we use, several limitations appear in \mo{} mock catalogues
because they are built from the replication of finite information. The
purpose of this section is to understand how the replication process
affects our predictions.

The most important limitations of \mo{} result from the fact that we
use a finite volume to describe the whole Universe. In order to do so,
we have to replicate the simulated volume many times along and across
the line of sight. Now, because we use the random tiling method to
proceed with these replications, some clustering information is
lost. This results in a negative {\it random tiling bias} which is
discussed in Sec. \ref{sec:rtb}. A more subtle effect comes from the
fact that the finite volume of the simulation used to build a mock
catalogue does not describe density fluctuations on large scales. Thus
these fluctuations will be missing from the mock catalogues. This
results in biases on counts variance estimates and correlation
function estimates. These {\it finite volume effects} are described in
Sec. \ref{sec:fve}.

In this section, we also check that other possible effects are under
control. In Sec. \ref{sec:fte}, we investigate the effect of finite
timestep on the apparent properties of galaxies in mock catalogues. In
Sec. \ref{sec:mre}, we check the impact of mass resolution of the root
simulation on different observable statistics.

\subsection{Random tiling bias} \label{sec:rtb}
A negative bias on correlation functions is introduced in mock
catalogues by the {\it random} tiling approach, which comes from the
fact that we decorrelate pairs of galaxies from one underlying box to
the other when re-shuffling them to suppress (periodic) replication
effects. Here, we first estimate this bias on the spatial two-point
correlation function, and then project the results to derive the bias
on the angular correlation function.

\subsubsection{Spatial correlation function (SCF)}
The spatial correlation function (SCF) can computed by
measuring the number of pairs of objects separated by a given
distance. If one uses the estimator of \citet[][hereafter LS93]{LandySzalay93} :
\beq
\xi(r) = \frac{DD(r) - 2 DR(r) + RR(r)}{RR(r)},
\eeq
where $DD(r)$, $DR(r)$ and $RR(r)$ are the number of data-data,
data-random and random-random pairs separated by distance $r$
respectively, one gets, through the logarithmic differentiation, the
relative error on $\xi(r)$ :
\beq
\frac{\delta \xi(r)}{\xi(r)} = \frac{\delta DD(r)}{DD(r)},
\eeq
because the random sample is not affected by the replication
technique.

Now, take a galaxy near the edge of an underlying box (say at a
distance $d < r$ from the edge). The mean number of pairs that will be
missed for this galaxy, due to replication, is :
\beq
\delta \widetilde{DD}_g(d,r) = 2\pi r^2 n \xi(r) (1-d/r) \dd r,
\eeq
where $n$ is the mean density of galaxies, and the subscript $g$
denotes that $DD$ is the number of pairs lost for {\it one}
galaxy. The tilde over $DD$ denotes that we only consider pair loss
through one side of the box. To compute the loss of pairs due to one
side of the underlying box, we integrate the previous equation over $d$,
from $0$ to $2r$, namely :
\beq\label{eq:perte par surface}
\nonumber
\delta \widetilde{DD}(r) &=& 2\pi r^2 n \xi(r) \dd r \int_{d=0}^{d=2r} n L_{box}^2 (1-d/r) \dd d \\
                 &=& 2\pi r^3 L_{box}^2 n^2 \xi(r) \dd r.
\eeq
Note that we
neglected corner effects here. This is justified by the fact that one
should always consider separations much smaller than the size of the
box (i.e. $r \ll L_{box}$).

To get the pair loss over a whole box, simply multiply the previous
result by 6 (the number of sides) :
\beq
\delta DD(r) = 12\pi r^3 L_{box}^2 n^2 \xi(r) \dd r.
\eeq

Had we not broken the continuity in the density field between each
underlying box, the number of pairs would simply be, for a whole box :
\beq \label{eq:nb pairs}
DD(r) = 4\pi r^2 \dd r n(1+\xi(r)) \times n L_{box}^3,
\eeq
where the first right hand term is the number of pairs expected for
one galaxy, and the second right hand term is the number of pairs in
the box.  The relative error in the number of pairs, due to shifting
the underlying boxes is thus :
\beq \label{eq:error xi}
\frac{\delta DD(r)}{DD(r)} = 3 \frac{r}{L_{box}} \frac{\xi(r)}{1+\xi(r)}.
\eeq

For a numerical estimate, consider the Lyman break galaxies (LBG)
population. For these galaxies, we expect $\xi(r) = 1$ at $r_0 \sim 6\
h^{-1}$Mpc. Thus, for our simulation, with $L_{box} = 100\ h^{-1}$Mpc,
one finds $\delta\xi / \xi < 10\%$ for LBGs, at $6\ h^{-1}$Mpc.

\begin{figure}
\centerline{\psfig{figure=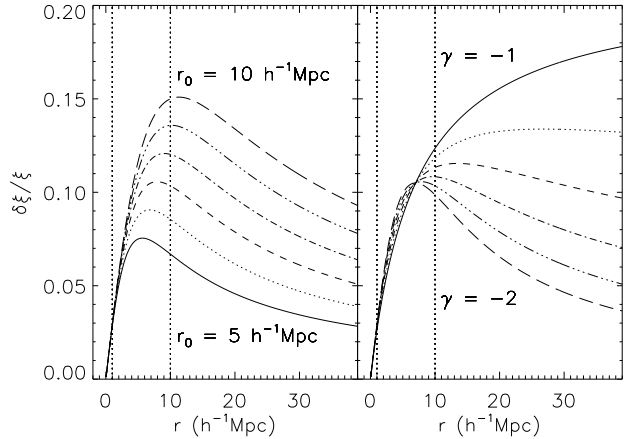,width=\hssize}}
\caption{Expected relative underestimation on the spatial correlation
function, assuming $\xi(r) = (r/r_0)^{\gamma}$. The left hand side
panel shows variations with $r_0$, with $r_0 = 5, 6, 7, 8, 9, 10$
$h^{-1}$Mpc from the bottom curve to the top curve, $\gamma$ being
fixed to $-1.8$. The right hand side panel shows the dependence on
$\gamma$, with $\gamma$ spanning the range [-2,-1], from bottom to
top, $r_0$ being fixed to $7 h^{-1}$Mpc. On each panel, the right hand
side vertical line shows the approximate upper limit of validity of
measurements of $\xi$ (one tenth of $L_{box}$). The left hand side
vertical line roughly indicates the size of a cluster, below which our
spatial information is uncertain.}
\label{fig:erreur_xi}
\end{figure}

On figure \ref{fig:erreur_xi}, we show the theoretical underestimation
on spatial correlation function measurements from our catalogues. In
the plots, we assume a correlation of the form
$\xi(r)=(r/r_0)^{\gamma}$, and we let $r_0$ and $\gamma$ vary. For a
wide range of these two parameters, the error due to transforming the
underlying boxes is less than 10\% from 1 to 10 $h^{-1}$ Mpc.

\begin{figure}
\centerline{\psfig{figure=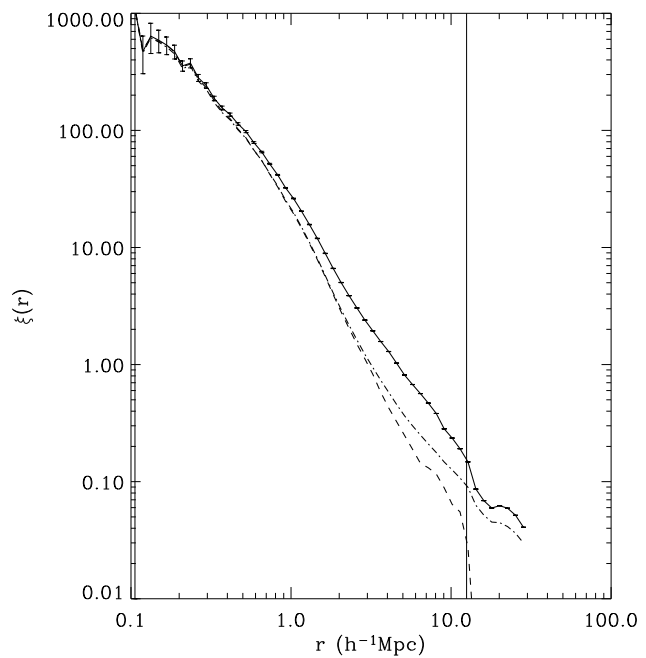,width=\hssize}}
\caption{Measured spatial correlation function of the dark matter
particles in the $z=0$ snapshot (solid curve) and on the same snapshot
re-shuffled (dashed line). Also plotted is the prediction from
Eq. \ref{eq:error xi} (dot-dashed line). The error bars attached to
the solid line are Poissonian errors. The vertical line indicates the
size of sub-boxes used for the re-shuffling.}
\label{fig:erreur_xi_num}
\end{figure}

On figure \ref{fig:erreur_xi_num}, we show a measure of the bias on
$\xi$ introduced by randomising the boxes. To do this, we cut our
snapshot volume into $8^3$ sub-boxes to which we applied translations,
rotations and inversions as described above. We then measured $\xi(r)$
on the original snapshot (solid curve) and on the shuffled
snapshot (dashed line). We plotted on figure
\ref{fig:erreur_xi_num} the prediction for the bias according to the
above calculation as the dot-dashed line. The agreement between
the measurement and the analytical prediction is very good on scales
up to $\sim 1/5 L_{box}$ (with $L_{box}$ being $100/8 h^{-1}$Mpc, as
shown by the vertical line on figure \ref{fig:erreur_xi_num}). In the
previous section, we showed that this scale is where finite volume
effects come into play. Although most finite volume effects are not
present here because we use all the sub-boxes to fill the simulated
volume, the randomisation of sub-boxes kills any signal at scales
larger than a sub-box. This is not described by the above analytic
calculation and results in the sharp cutoff of $\xi$ at $\sim 1/5
L_{box}$.

\subsubsection{Angular correlation function (ACF)}
We can use the bias on the SCF derived above to evaluate that on the
ACF. Let's first remember that, in the small angle approximation, the angular correlation is related to the spatial one by \citep{BernardeauEtal02},
\beq
\label{eq:bernardino}
w(\theta) = \int \dd \chi \ \chi^4 D_M(\chi) \psi^2(\chi) \int_{-\infty}^{\infty} \theta \xi(r) \dd x,
\eeq
where $\chi$ is the radial distance, $D_M$ is the angular distance
($D_M=\chi$ in a flat universe), $\psi$ is a selection function
satisfying $\int \chi^2 \psi(\chi) \dd \chi = 1$, and $r$ is the
separation distance, related to the angular separation $\theta$ and
the integration variable $x$ through the relation $r = D_M \theta
(1+x^2)^{1/2}$. 
If we now introduce the bias on $\xi$ as $\xi \mapsto \xi +
\delta\xi$, with $\delta\xi$ given by Eq. (\ref{eq:error xi}), we
can derive the corresponding bias on the ACF from Eq. \ref{eq:bernardino}~:
\beq
\delta w(\theta) = \int \dd \chi \ \chi^4 D_M(\chi) \psi^2(\chi) \int_{-\infty}^{\infty} 3 \theta \frac{r}{L_{box}} \frac{\xi^2(r)}{1+\xi(r)}\dd x
\eeq
Assuming that $\xi$ can be written as the power law
$(r/r_0)^{-\gamma}$, and using $D_M(\chi)=\chi$ in a flat universe,
we end up with :
\beq \label{eq:delta_w}
\delta w(\theta) &=& \frac{3r_0^{\gamma}\theta^{2-\gamma}}{L_{box}} \\ \nonumber
	&\times& \int_0^{\infty} \dd \chi \chi^{6-\gamma} \psi^2(\chi)
	\int_{-\infty}^{\infty} \frac{(1+x^2)^{1/2 - \gamma} \dd x }
	{\left(\frac{\chi \theta}{r_0}\right)^{\gamma} + (1+x^2)^{-\gamma/2}}
\eeq

Note that for this result, we also assumed that the SCF does not vary
with redshift. This is obviously wrong in general but is justified if
the selection function $\psi$ is narrow enough (e.g. for LBGs).

\begin{figure}
\centerline{\psfig{figure=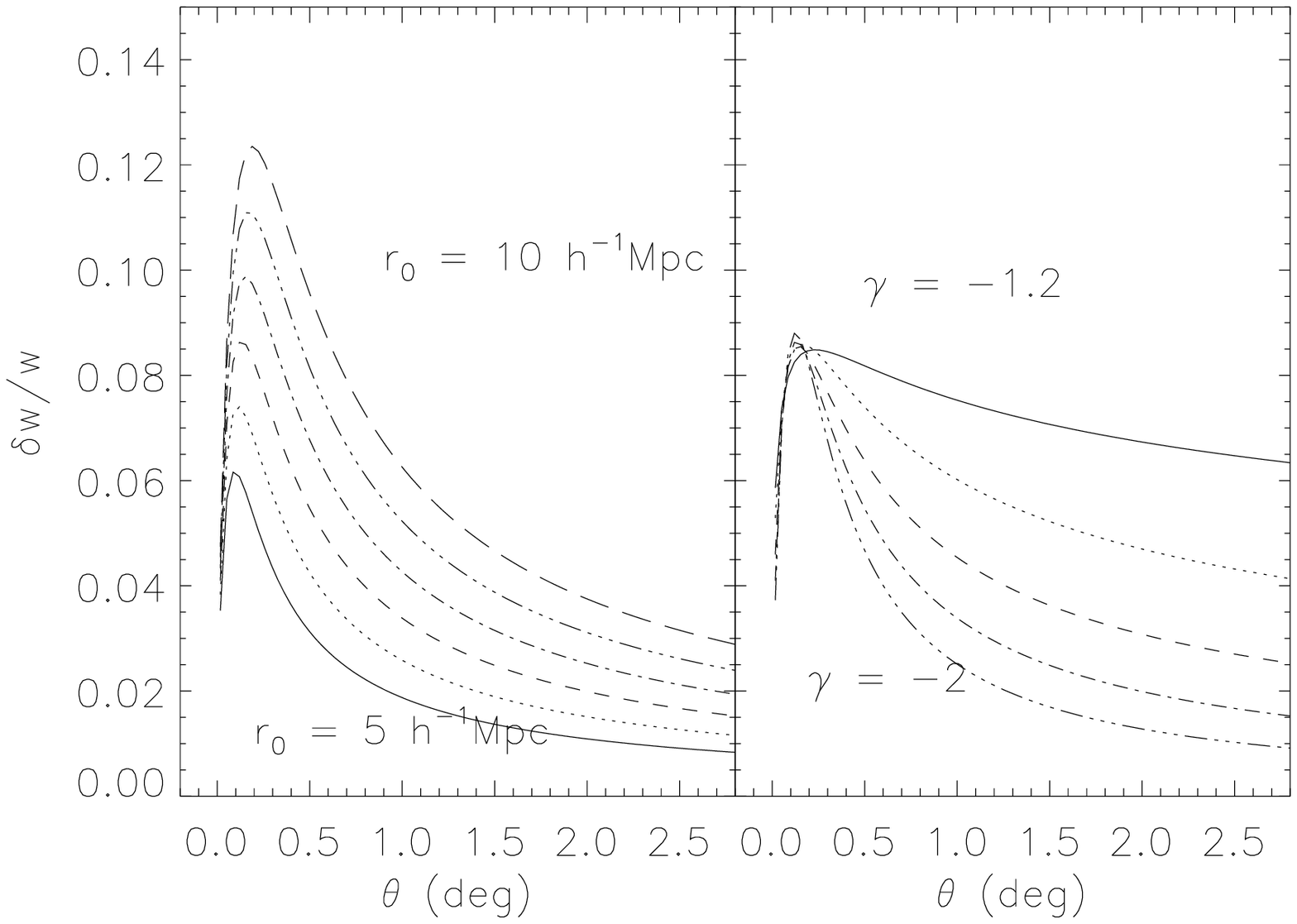,width=\hssize}}
\caption{Expected relative underestimation on the angular correlation
function, assuming $\xi(r) = (r/r_0)^{\gamma}$. The left hand side
panel shows variations with $r_0$, with $r_0 = 5, 6, 7, 8, 9, 10$
$h^{-1}$Mpc from the bottom curve to the top curve, $\gamma$ being
fixed to $-1.8$. The right hand side panel shows the dependence on
$\gamma$, with $\gamma$ spanning the range [-2,-1], from bottom to
top, $r_0$ being fixed to $7h^{-1}$Mpc. The selection function chosen
here is simply a top hat centred at $\chi=2000h^{-1}$Mpc and of width
500$h^{-1}$Mpc.}
\label{fig:erreur_w}
\end{figure}

Finally, using equations (\ref{eq:bernardino}) and (\ref{eq:delta_w}), and
deciding on a selection function, one can compute numerically the
relative bias induced on ACF measurements by the transformations of
underlying boxes. An example is given in figure \ref{fig:erreur_w}.

\subsection{Finite volume effects} \label{sec:fve}
Several limitations arise because we use a finite volume to describe
the whole Universe. They are basically due to the fact that a finite
volume $V$ does not describe density fluctuations on scales typically
larger than $\sim V^{1/3}$. In other words, although the mean number
of galaxies in a simulation can be tuned to fit that observed in the
Universe, the simulation does not describe the dispersion about this
mean value. How this affects statistics from our catalogues is the
question we address in this section. The simplest statistic we are
interested in is galaxy counts, as a function of magnitude or
redshift. Mock catalogues can be used in two ways : (i) to normalise
models, and (ii) to estimate errors (including cosmic variance). In
Sec. \ref{sec:counts}, we discuss how finite volume affects both
the counts and their variance. Then, in
Sec. \ref{sec:finite_clustering}, we describe the bias on correlation
functions introduced by finite volume effects.

\begin{table*} 
\begin{tabular}{l c c c c c c}
\hline \hline
Sample & Area & Selection criteria & $<z>$ ($z_{min}$ ; $z_{max}$) & $<i>$ ($i_{min}$ ; $i_{max}$) & $\theta_{100}$ (deg)& $N_{100}$ \\
\hline
APM    & 100 deg$^2$ & $17 < B_J < 20$    & 0.18 (0.04 ; 0.73)            & 65.3 (54 ; 69)                & $\sim 10$           & $\sim 5.5$ \\
K20    & 1 deg$^2$   & $K_s < 20$         & 0.53 (0.02 ; 1.74)            & 58.0 (41 ; 69)                & $\sim 4.5$          & $\sim 12$  \\
Counts & 1 deg$^2$   & $\chi<5500h^{-1}$Mpc&           -                   &          -                    &    -                &      -     \\ 
\hline
\end{tabular}
\caption{Geometry of mock catalogues used for comparison to data from
the APM and the K20 (see text), and to make $K_s$-band counts. All mock catalogues have a square surface, of area given in the second column. The third column states how galaxies are selected in these catalogues (in the ``Counts'' case, no photometric selection is applied, but the catalogues are truncated at a comoving distance of $5500h^{-1}$Mpc from the observer). The fourth column gives the mean and span of the redshift distribution. The fifth column gives the mean and range of outputs used (output 70 is $z=0$). The sixth column gives the angular size of the simulated volume (of side $L_b=100h^{-1}$Mpc) at the mean redshift of the sample. The last column gives the number of radial replications needed to reach the mean redshift using the full simulated volume.\label{tab:samples}}
\end{table*}

\subsubsection{Effects on estimates of counts variance} \label{sec:counts}

Two variances are relevant for counts in mock
catalogues. The first is the variance which tells us about the dispersion
of number counts from mock catalogues each generated from a {\it
different} simulated volume. The second is the variance that describes
the dispersion in number counts from mock catalogues made from a {\it
unique} simulated volume. This variance tells us to what extent we can
estimate cosmic variance with mock catalogues based on a given
simulation. Let us proceed to virtual experiments to understand these quantities.

\begin{itemize}
\item Imagine we have a large number $N$ of simulations at hand, all describing an equal volume $V$, but with initial conditions drawn from a much larger volume. From each simulation, we build a mock catalogue using \mo{}, and then count galaxies brighter than some magnitude limit. Finally, we measure the variance $\sigma_1^2$ of the counts obtained in this way. Now, imagine that we also have $N$ mock catalogues, each generated using an ideal technique and a simulated volume much larger than that of the light-cone. Call $\sigma_2^2$ the variance in the counts measured from these catalogues. In the case where the volume of the cone is much smaller than volume $V$, the two above variances will be equal. In the more realistic case where $V$ is smaller than the volume of the cone, one will measure that $\sigma_1 > \sigma_2$ : replication enhances the bias of the simulated volume, thus dispersing more counts from catalogues.

\item As a second experiment, imagine one has a unique simulated volume $V$ as above, and builds many mock catalogues from it. These catalogues will be different from one another because of the random tiling process and because the light-cone may intersect different sections of $V$ in different realisations. As before, measure the variance $\sigma_3^2$ of the counts from these catalogues. Again, if the simulated volume is much larger than that of the light-cone, one will measure $\sigma_3^2 \sim \sigma_2^2$. In this case, one can use mock catalogues to estimate cosmic variance. However, in the case where the volume of the cone is larger than $V$, $\sigma_3$ will be found lower than $\sigma_2$, because replications do not add large-scale fluctuations. In the extreme case where the cone is very large compared to the simulated volume, $\sigma_3$ will tend to zero, because the cone encloses all the information contained in $V$.
\end{itemize}

\begin{figure}
\psfig{figure=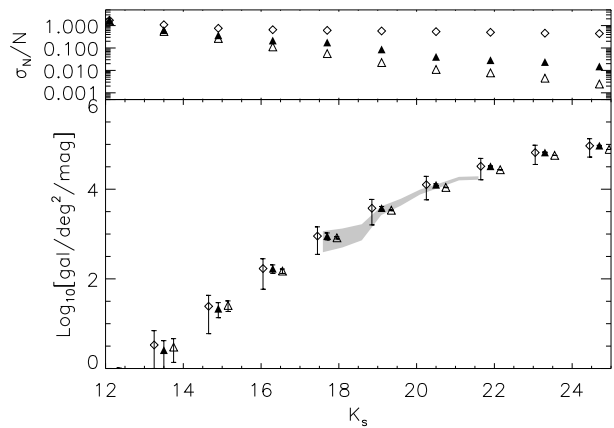,width=\hssize}
\caption{Effect of finite volume on $K$-band counts. {\it Lower
panel}~: the filled (respectively open) triangles show the mean number
counts obtained from 20 catalogues made from our standard simulation
(resp. from one sub-box), the error bars giving the standard
deviation about this mean. The open diamonds show the mean number
counts measured from 125 mock catalogues, each built from a different
sub-box. The shaded area shows the locus of the K20 counts from
\citet{CimattiEtal02a}. {\it Upper panel} : relative standard deviation of the counts
(same symbol code as lower panel).}
\label{fig:dbs_counts}
\end{figure}

In Fig. \ref{fig:dbs_counts}, we show K-band counts measured from
various mock catalogues having the geometry defined in the third line
of Table \ref{tab:samples} (``counts'' catalogues). The shaded area
shows the locus of K20 counts from \citet{CimattiEtal02a}, including
Poissonian error bars. The filled symbols and their error bars give
the mean and standard deviation for counts measured from 20 mock
light-cones made using the standard simulated volume. Then we cut our
root simulation into 125 sub-boxes, and made a mock catalogue out of
each sub-box.  The open diamonds give the mean and standard deviation
of counts measured from these 125 mock light-cones. Finally the open
triangles give the mean and standard deviation of counts measured from
20 mock catalogues made from a single sub-box. The upper panel of
Fig. \ref{fig:dbs_counts} compares the relative standard deviations of
these three measures.

First, note that changing the size of the volume used to make mock
catalogues does not change the shape of the counts, but only the
amplitude. This was expected since evolution is the same in all
sub-boxes. This tells us that if our root simulation is well
normalised, counts from mock catalogues are not sensitive to the size
of the simulated volume and can thus be used with confidence to normalise theory to observations.

Second, consider the difference between open and filled triangles in
the upper panel of Fig. \ref{fig:dbs_counts}, that is, estimates of
cosmic variance from cones made using one sub-box or the full
simulated volume. As expected, we find that using a small box leads to
an under-estimate of the variance. Table \ref{tab:samples} tells us
that the number of full boxes used to describe galaxies brighter than
$K_s=20$ is about 12 along the line of sight (up to the median
redshift) and $1/5$ across the line of sight at the median
redshift. This situation is at the limit where we can correctly
estimate cosmic variance, since the light-cone only intersects a
fraction of the box volume in each underlying box. For the sub-box
case, the light-cones include a full underlying box at the median
redshift, and about 60 sub-boxes are replicated along the line of
sight to reach this redshift. In this regime, the angular correlation
function is largely under-estimated at the scale of the catalogue
(because it is larger than the scale of a box), and so the estimated
cosmic variance is under-estimated too.

Third, it is interesting to consider the difference between the filled
triangles and the open diamonds. At bright magnitudes, the two give the
same variance. This is because the volume probed by the mock
catalogues is much smaller than the volume of a sub-box, so variance
is well estimated with both methods, and is in fact Poissonian. At
intermediate magnitudes, the volume probed by the mock cone is smaller
than the full simulated volume, yet larger than that of a
sub-box. Hence, the sub-box variance saturates at higher values. At the faint end,
the light-cone is larger than the full simulated volume, so the
variance showed with filled triangles suffers from a similar negative
bias as that shown by the open triangles. The three regimes are
spanned here and show that in practice, robust estimates of cosmic
variance require a simulated volume much larger than
the volume probed by the mock catalogue.

\begin{figure}
\psfig{figure=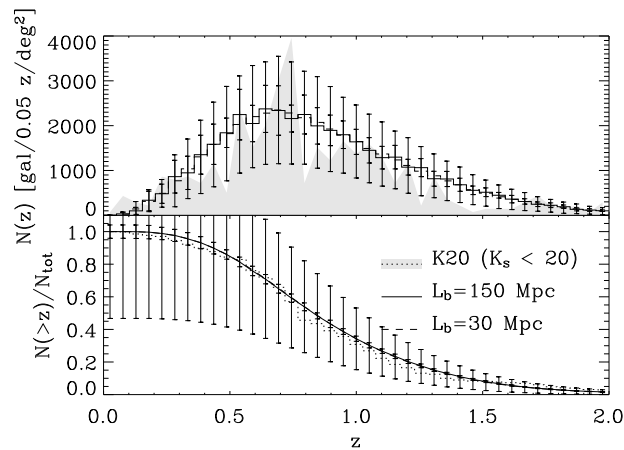,width=\hssize}
\caption{Effect of finite volume on the redshift distribution of a
flux-limited sample. The solid (resp. dashed) curve shows the mean
redshift distribution estimated from 20 mock catalogues made from the
full simulated volume (resp. 125 mock catalogues each made from one
different sub-box). The attached error bars give the standard
deviations (the small ones correspond to the  solid curve). The upper panel gives the differential distributions while
the lower panel gives the normalised cumulative distributions. The
shaded area and dotted line show the locus of the K20 data in the top and bottom panels (see text). The agreement
with observations is very good.}
\label{fig:dbs_nz}
\end{figure}

Redshift distributions are affected by finite volume effects in two
ways. First, the variance and mean of the redshift distributions will
change with box size. This effect is the same as that described above
for the counts. Second, because the smaller the box, the more
replications involved, repeated structures may imprint periodic
features in $N(z)$. Thanks to the random tiling technique, this
problem is avoided. In Fig. \ref{fig:dbs_nz}, we show the differential
(upper panel) and cumulative (lower panel) redshift distributions of
galaxies selected as in the K20 survey \citep{CimattiEtal02}. The shaded area (resp. dotted curve) indicates the
locus of the data in the upper panel (resp. lower panel). In both panels, the curves (which are mostly
over-imposed) show the mean distributions measured from 20 mock
catalogues made with the full simulated volume and from 125 mock
catalogues each made with a different sub-box. The error bars show the
standard deviations about these means (the larger error bars
correspond to the sub-box catalogues). Fig. \ref{fig:dbs_nz} shows
that the redshift distribution does not change in shape when the size
of the box varies. Although the details of $N(z)$ will differ from one
catalogue to the other, the statistical significance of redshift
distributions is thus robust, and found to be in good agreement with $K$-band
observations. 

\begin{figure}
\psfig{figure=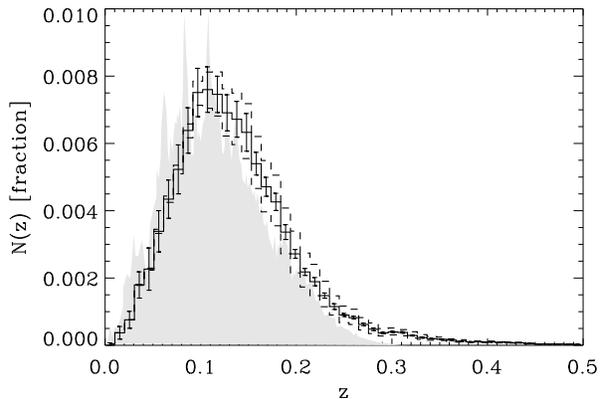,width=\hssize}
\caption{Comparison of \gal{} to the 2dFGRS redshift distribution. The shaded shows data from \citet{CollessEtal01}. The solid histogram (resp. lower and upper dashed histograms) shows the \gal{} redshift distribution for galaxies brighter than $b_J=19.3$ (resp. 19.2, 19.4) estimated from 20 mock surveys of 10$\times$75 square degrees. The error bars show the dispersion in the estimates. The same arbitrary normalisation was applied to the 3 histograms.}
\label{fig:2dFnz}
\end{figure}

The agreement found with the K20 redshift distribution is an important
success of the \gal{} model, given the difficulty experienced by other
models in achieving this task. The redshift distribution observed in
the 2dFGRS \citep[e.g.][]{CollessEtal01} also seems to have been
challenging for modellers to reproduce, and it is interesting to see
how \gal{} and \mo{} pass this test. The shaded area in
Fig. \ref{fig:2dFnz} shows the redshift distribution of 2dF galaxies
given by \citet{CollessEtal01}. This distribution includes the whole
survey, and thus corresponds to a {\it nominal} magnitude cut at
$b_J=19.45$. Because of various sources of incompleteness, however,
the {\it effective} magnitude cut is more likely to lie around $b_J =
19.3$ \citep[see Fig. 14 from][]{CollessEtal01}. The solid histogram
in Fig. \ref{fig:2dFnz} shows the average redshift distribution
measured in 20 mock surveys of 10$\times$75 square degrees, limited in
apparent magnitude at $b_J=19.3$. The associated error bars show the
dispersion around this mean. The dashed lines show the redshift
distributions corresponding to an apparent magnitude cut at $b_J=19.2$
(lower line) and $b_J=19.4$ (upper line). The comparison suggests that
the evolution of the $b_J$-band luminosity function predicted by
\gal{} is incorrect, giving too many bright galaxies at high
redshifts. Indeed, an apparent magnitude cut at $b_J\sim 19.1$ is
necessary to bring our redshift distribution into better agreement
with the 2dFGRS results. Let's note however that the scope
of this comparison is limited in several ways. First, one should
include a proper description of the complex selection function of the
2dFGRS for a more meaningful comparison. Although beyond the scope of
this paper, this is readily feasible by applying the masks of the
2dFGRS to \mo{} mocks with similar geometry. Second, the dispersion
showed in Fig. \ref{fig:2dFnz} tells us that despite the huge amount
of data gathered by the 2dFGRS, cosmic variance is still quite
large. Following the above discussion on finite volume effects, and
looking back at Fig. \ref{fig:ex_light_cone}, one sees that this
dispersion is bound to be an under-estimate of the true cosmic
variance because of the many replications involved in 2dFGRS-like mock
surveys. This tells us that we need a bigger simulated volume to
actually constrain the model~: one needs to have realistic cosmic
variance at a survey's size before hoping to discriminate between
different models. Finally, this example shows how useful the \mo{}
software is to carry out detailed comparisons of models with various
datasets.

\subsubsection{Effects on estimates of 2-point correlation functions} \label{sec:finite_clustering}
\begin{figure}
\psfig{figure=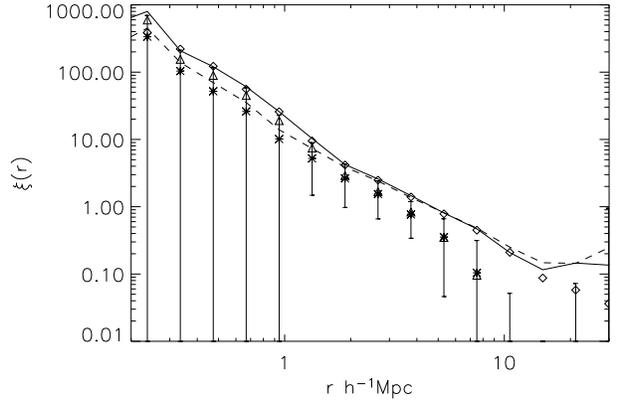,width=\hssize}
\caption{Spatial correlation function of galaxies brighter than
$B=-19$. The diamonds show $\xi_{100}$, computed from the whole
simulation, and the stars show the mean of estimates of $\xi_{20}$
from 125 sub-boxes of side $L_{sb}=20h^{-1}$Mpc (stars). The error
bars show the standard deviation about this mean. The dashed line
shows $\xi_{20}$ corrected for finite volume effects (see text), and
agrees with the diamonds at large separations.}
\label{fig:SCF_finite}
\end{figure}

Finite volume effects alter correlation functions in a complex
way. Let's first discuss what happens to the spatial correlation
function (SCF) in a cubical box such as the simulated volume. The
situation in mock catalogues is analogous but also includes projection
effects.  Following LS93 we relate the correlation function
$\hat{\xi}$ contained in the simulated volume $V$ to the ``real''
$\xi$ as
\beq \label{eq:IC_LS} 
1+\hat{\xi} =\frac{1+\xi}{1+\bar{\xi}_V}.  
\eeq 
At small separations (compared to $V^{1/3}$), where $\bar{\xi}\ll
\xi$, the bias is negligible. At large scales, $\bar{\xi}\sim \xi$ and
$\hat{\xi}$ falls down to 0. This bias directly results from the fact
that the variance cannot be estimated properly at the simulated volume
scale, from only one simulated volume. Let's carry out
numerical tests to better understand the finite volume bias on the spatial
correlation function. We again cut our standard simulation
of side $L_b=100h^{-1}$Mpc into 125 cubic sub-boxes of side $L_{sb} =
20h^{-1}$Mpc, and we measure the spatial correlation function (SCF) in
all these 126 boxes, for galaxies brighter than $B=-19$. This
magnitude cut leaves us with about 150 galaxies per sub-box. In
Fig. \ref{fig:SCF_finite}, we plot the SCF measured from the full
simulation ($\xi_{100}$) with diamonds, and the average of the 125
measures on sub-boxes ($\langle\xi_{20}\rangle =\sum\xi_{20}/125$) as
stars. The error bars attached to the stars show the standard
deviation from the 125 estimates of $\xi_{20}$. Comparison of
$\xi_{100}$ and $\langle\xi_{20}\rangle$ shows that finite volume
effects translate into a negative bias at all scales, with a rather
sharp cutoff at $r\sim L_{sb}/5$. The dashed line shows $\langle
\xi_{20} \rangle$ corrected from the integral constraint given in
Eq. \ref{eq:IC_LS}. The agreement of the dashed line with the diamonds
is very good at large scales. At separations smaller than $r\sim
1h^{-1}$Mpc, sub-box--to--sub-box fluctuations (both due to sparse
sampling and to clustering) are responsible for the remnant
discreteness bias. To understand this, let's consider the
pair-weighted average of $\xi_{20}$~:
\beq 
\tilde{\xi}_{20} = \frac{\sum_i n_i^2\xi_{20,i}}{\sum_i n_i^2}, 
\eeq 
where $\xi_{20,i}$ is the correlation function measured (with the
estimator from LS93) on the $n_i$ galaxies of sub-box $i$. This
weighted average is shown with the triangles on
Fig. \ref{fig:SCF_finite}.  Notice that in the case where edge effects
are negligible (i.e. at small separations), one finds
\beq
\tilde{\xi}_{20} \simeq \frac{\sum_i DD_i - 2 DR_i + RR_i}{\sum_i RR_i} \simeq 
\frac{DD - 2 DR + RR}{RR},
\eeq
where $DD$, $DR$, and $RR$ are the numbers of data-data, data-random,
and random-random pairs in a given separation bin for the whole
simulated box. In other words, the pair-weighted average of the
correlation functions of the sub-boxes is equivalent, at small scales,
to using the LS93 estimator for the whole box, and is thus only affected
by the integral constraint. Now, the main difference between this
estimate and the estimate obtained from the full simulation (open
diamonds) is that cross pairs between two sub-boxes are not
regarded. In particular, as expected, this estimator converges to the
biased estimator $\langle\xi_{20}\rangle$ at large scales. But at
small scales, $\tilde{\xi}_{20}$ partly captures sub-box--to--sub-box
fluctuations through the variations of $n_i^2$, and thus remains above
$\langle\xi_{20}\rangle$. Still, $\tilde{\xi}_{20}$ is a combination
of estimates of the SCF on small boxes, which are contaminated at all
scales by the integral constraint effect. Hence $\tilde{\xi}_{20}$
remains below the ``exact'' result. When $\tilde{\xi}_{20}$ is
corrected from the integral constraint as in Eq. \ref{eq:IC_LS} (solid
line in Fig. \ref{fig:SCF_finite}), the result matches nearly
perfectly $\xi_{100}$, as expected.

\begin{figure}
\psfig{figure=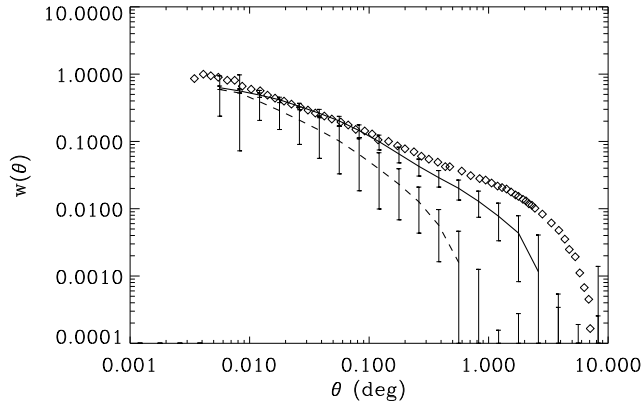,width=\hssize}
\caption{The solid (resp. dashed) curve show the mean ACF estimated
from 20 mock catalogues made from the full simulated volume (resp. 125
mock catalogues each made from one different sub-box). The error bars
give the standard deviations about these means. The cutoff in the
sub-box-catalogue ACF occurs at about a fifth of the angular size of a
sub-box taken at the mean redshift of the galaxy sample. Open diamonds
show the ACF measured from the APM
\citep{MaddoxEfstathiouSutherland96}.}
\label{fig:dbs_w}
\end{figure}

Fig. \ref{fig:dbs_w} shows how these finite volume effects affect the angular
correlation function. The solid line shows the mean of angular
correlation functions measured from 20 mock APM catalogues (see Tab. \ref{tab:samples}) made from
the full simulated volume, the attached error bars give the measured
standard deviation. The dashed line and corresponding error bars show
the mean and standard deviation of the ACF measured from the 125 mock
APM catalogues made from the sub-boxes.

For the dashed line, the departure from a power-law at large scales
reported in the 3-D case (see Fig. \ref{fig:SCF_finite}) occurs here
at $\theta\sim 0.4$ degree. This is a direct consequence of the finite
volume of the sub-box, and 0.4 degree is here about one fifth of the
angular size of a sub-box at the median redshift of the survey (see
Table \ref{tab:samples}). On top of this turn-around, there is an
overall bias which increases slowly with separation starting at scales
of about one hundredth the size of a sub-box. This is due to the
projection of the bias described above for the SCF. Now, the open
diamonds on Fig. \ref{fig:dbs_w} show the ACF measured from the APM by
\citet{MaddoxEfstathiouSutherland96}. The data are in very good
agreement with our (full-box) model at scales shorter than $\sim 0.1$
degree -- which is about $L_b/100$ at the median redshift. Long-wards
of this scale, finite volume bias our ACF progressively. The comparison
of our full-box $w(\theta)$ to data from the APM is similar to the
above comparison between sub-box and full-box ACFs. We thus understand
that the large-scale disagreement between APM and our model is not
physical, but due to finite volume effects : APM data are drawn from an
even larger box : the Universe !

Finally, let us come back to the issue of counts' variance in mock
catalogues. Remembering that the variance of counts is basically given
by the average of the angular correlation over the survey, we now
clearly see how $\sigma_3$ of previous section was
under-estimated. And we understand that this under-estimation will
occur unless we use a simulated volume more than ten times larger than
the aperture of the light-cone at the redshift of interest.

\subsection{Other effects}

\subsubsection{Timestep} \label{sec:fte}

The fact that we use a finite number of outputs, typically spaced in time by 100 Myr, could affect mock catalogues. We argued in Sec. \ref{sec:app_props} that this was not expected because even though individual galaxies can undergo a dramatic evolution during a timestep, the average properties (and their dispersion) of the overall population evolves at a much slower pace. Nevertheless, we check this hypothesis in this section by comparing statistics from mock catalogues made using different timesteps of the same simulation.
Namely, we compare the counts, redshift distributions and ACFs obtained
with our reference mock catalogues to those obtained with catalogues
made using one snapshot out of ten\footnote{Note that in any case, the properties of the
galaxies were computed using all timesteps, which is
necessary in order to properly describes the physics at stake in
galaxy evolution \citep[see][]{HattonEtal03}.}.

\begin{figure*}
\begin{tabular}{ccc}
\psfig{figure=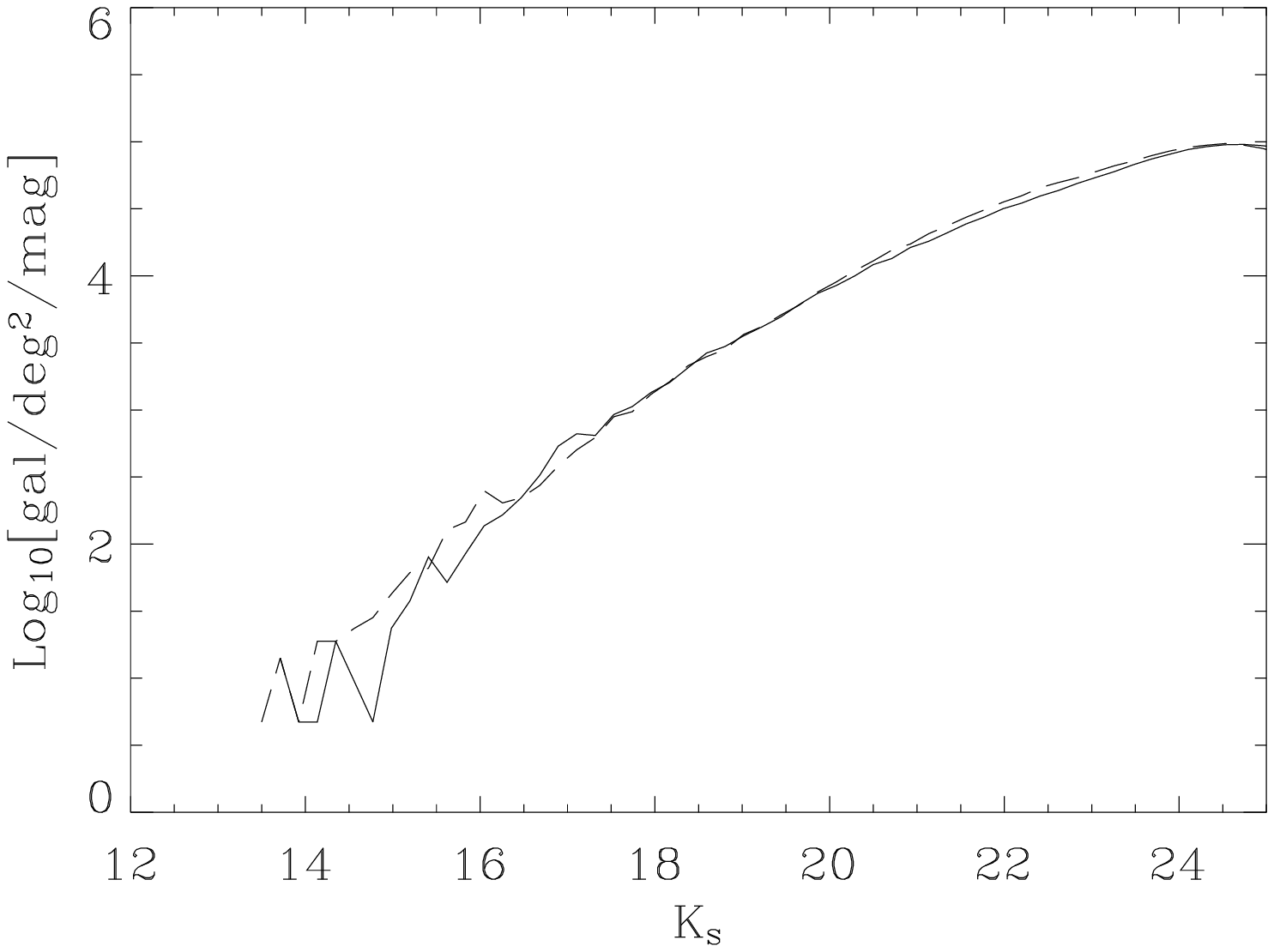,width=\htsize}
\psfig{figure=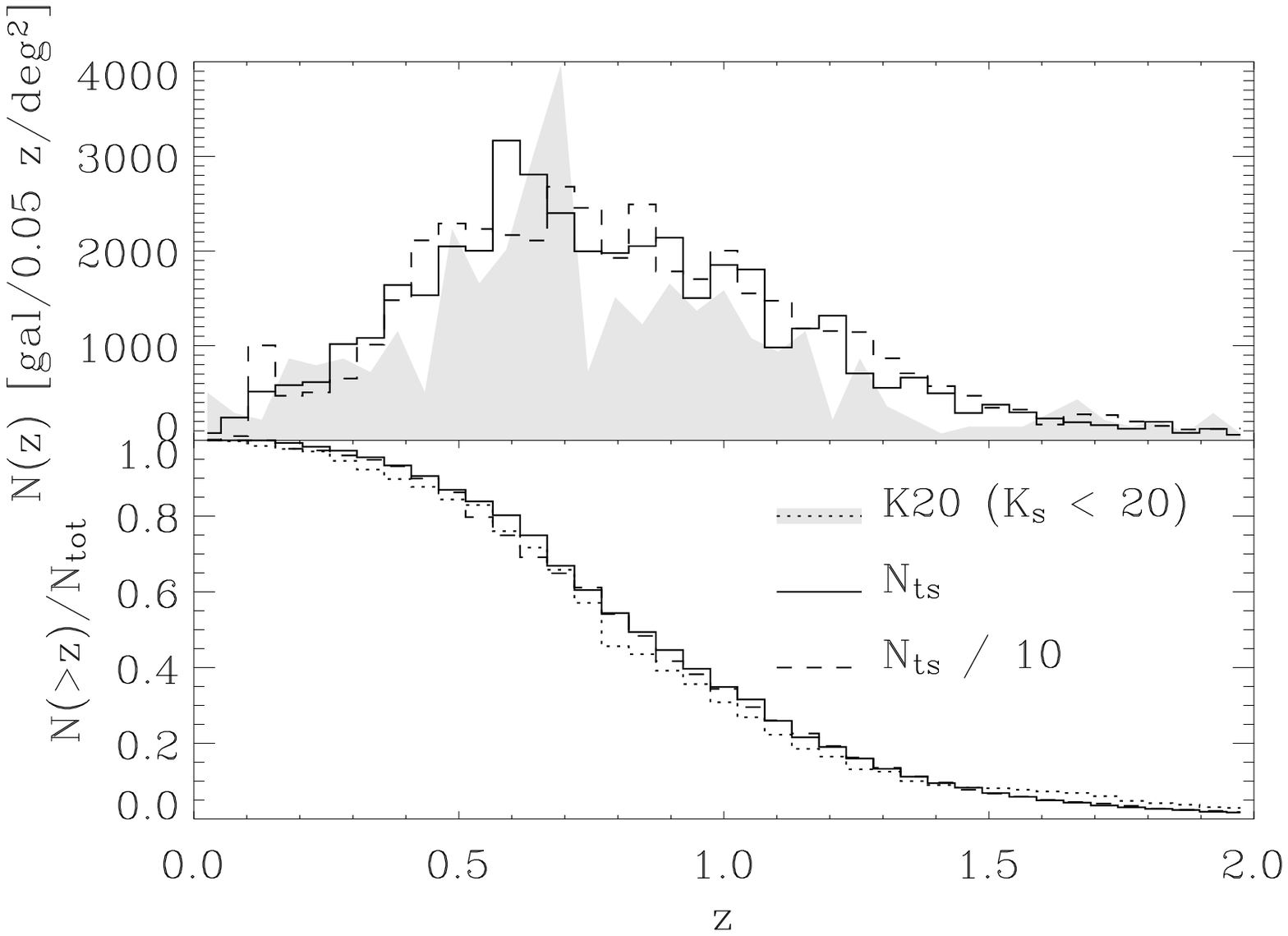,width=\htsize}
\psfig{figure=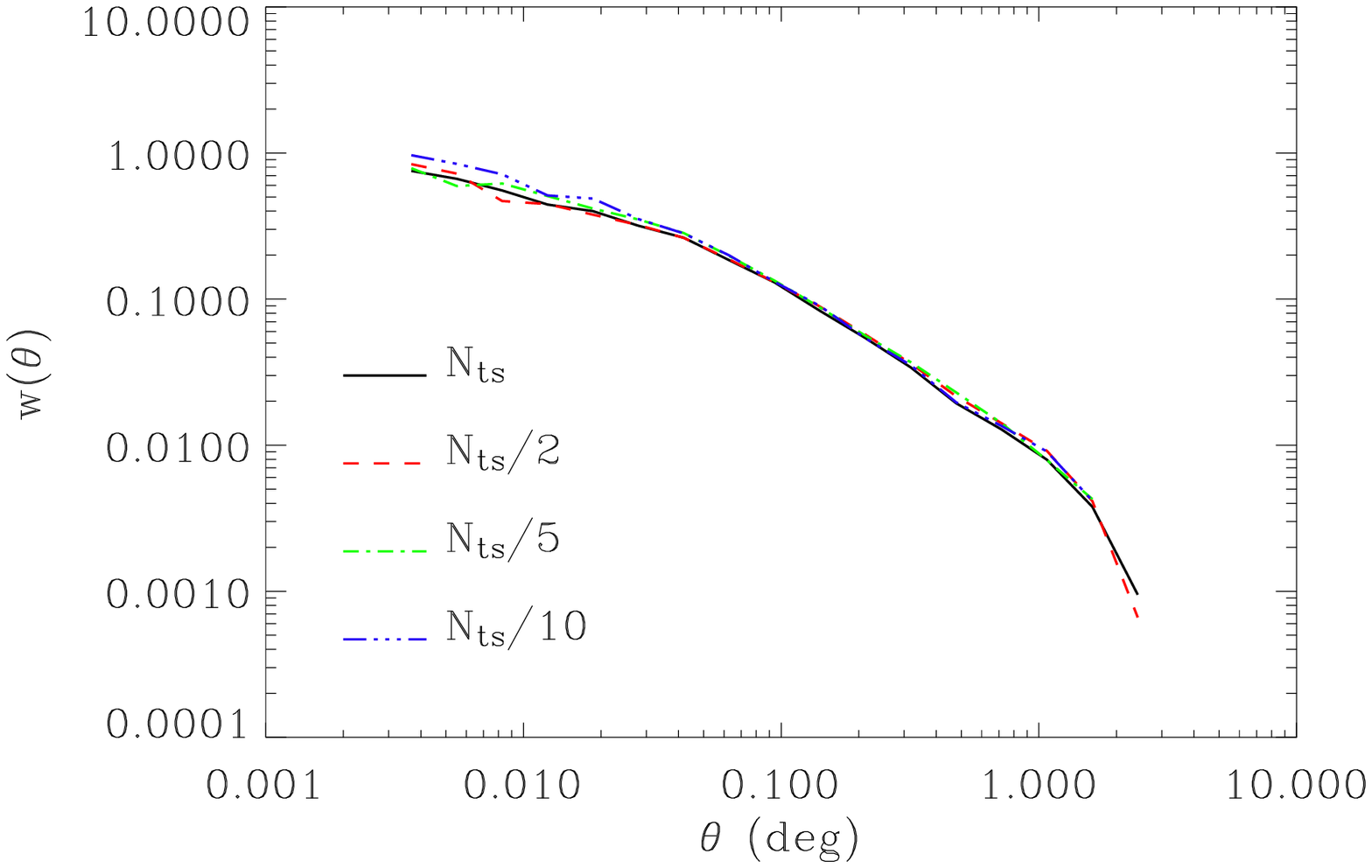,width=\htsize}
\end{tabular}
\caption{{\it Left hand side panel} : number counts from a mock catalogue using all available time-steps (solid curve) and one snapshot out of 10 (dashed curve). {\it Middle panel} : redshift distributions using all snapshots (solid histograms) or one out of 10 (dashed histogram). The data from the K20 survey are shown by the grey area in the upper panel and by the dotted line in the lower panel. {\it Right hand side panel} : angular correlation functions in catalogues containing all the snapshots (solid curve), one out of 2 (dashed curve), one out of 5 (dot-dashed curve) and one out of 10 (3-dot-dashed curve).}
\label{fig:dts}
\end{figure*}

The resulting counts, redshift distributions and ACFs are shown in
Fig. \ref{fig:dts}, and show no significant difference between the fine
and coarse time-steps. This shows that the random tiling method is
robust in that the resulting mock catalogues do not depend on the
time-step used in the root simulation, provided the physics was properly integrated. The fact that the properties of
mock catalogues do not change with time-step shows that, at least for
the selected galaxies, the K-correction and possibly a slow evolution
determine the statistics. This justifies a posteriori the first-order
correction made for magnitudes in Eq. \ref{eq:mag_correction}.

The necessity of using a fine time-step to make mock observations then
mainly comes from the complex analysis that can be made from them
\citep[see][]{BlaizotEtal04}. In this perspective, one wants to
retrieve the physical properties of individual galaxies, as well as
their hierarchical evolution, for samples selected according to
observational criteria. A short time-step naturally allows to analyse
evolution in more detail.

\subsubsection{Mass resolution} \label{sec:mre}

As mentioned in Sec. \ref{sec:resolution}, the mass resolution of the
DM simulation affects galaxies in three ways : (i) incompleteness,
(ii) limit redshift, and (iii) ``immaturity''. These three
limitations, inherent to the hybrid method implemented in \gal{}, will have different effects on statistics measured from mock catalogues.
\begin{itemize}

\item[(i)] {\it Incompleteness} sets in when a fraction of galaxies of a given sample are missed because they would lie in halos below the mass resolution of the DM simulation. This effect obviously causes under-estimates of the counts at faint magnitudes. A more subtle effect is that a sample of galaxies affected by incompleteness will have a halo mass distribution biased towards high masses. Because more massive halos are more clustered, this will induce a positive bias on correlation functions. These effects cannot be corrected for except by using simulations with better mass resolution. However, as shown in \citet{BlaizotEtal04}, it does not prevent one from using mock catalogues for studying bright galaxies, even at high redshift.

\item[(ii)] {\it The limit redshift} is the redshift beyond which no halo can be detected in a DM simulation. All possible galaxies at higher redshifts are thus missed by \gal{}, and are hence missing from our mock catalogues. This effect, combined with incompleteness is responsible for a faint-end decrease in the counts. 

\item[(iii)] {\it Immaturity} describes the fact that young galaxies have unrealistic properties mainly because the cooling of gas in their host haloes was not slowed down by DM accretion in sub-resolution progenitors. These galaxies only become a significant part of the overall population at redshifts higher than $\sim 2$ in our standard simulation, and they can be easily flagged and removed from a sample from our database. In terms of apparent magnitudes, they only significantly affect $K$ band counts at $K > 24$. 

\end{itemize}

The natural, and foreseen, solution to these limitations is to increase the resolution of root DM simulations. We come back to this perspective in the conclusions.

\section{Database and web interface}  \label{sec:database}
\begin{figure}
\psfig{figure=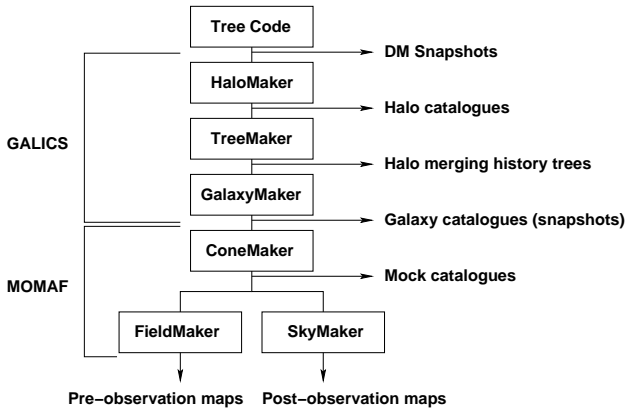,width=\hssize}
\caption{Organisation chart of the \gal{}/\mo{} project.}
\label{fig:schema_galics}
\end{figure}

The current implementation of the \gal~hybrid model of hierarchical
galaxy formation is a package that includes three main
routines (see Fig. \ref{fig:schema_galics}). First, {\tt HaloMaker} identifies haloes in each of the
output snapshots. Second, {\tt TreeMaker} constructs the halo merging history
trees from the list of halos in all the snapshots, and computes the
dark matter properties of each of the halos. Third, {\tt GalaxyMaker}
deals with the fate of baryons within the merging history trees. It
computes the properties of hot gas in halos, and follows galaxy
formation and evolution. The outputs are a list of properties (including absolute
magnitudes in standard photometric bands) and rest-frame spectra for
all galaxies in snapshots. The information produced by a given
\gal~post-processing of the simulation (defined by the choice of the
astrophysical free parameters) constitutes what we hereafter call the
{\it Archives} of this post-processing (see \hatton). Any change in
the list of input parameters will correspond to a new issue of the
post-processing and new output {\it Archives}.

The \mo{} post-processing is also a package with three main routines,
detailed in previous sections (see
Fig. \ref{fig:schema_galics}). First, {\tt ConeMaker} generates an
observing cone by integrating along the light-cone through the
various snapshots, and by managing the radial and transverse
replications. The output is a list of galaxies with apparent
properties (including apparent magnitudes in standard photometric
bands) that are computed from the {\it Archives}. Second, {\tt
FieldMaker} builds mock images from the observing cone, by projecting
the cone galaxies onto the plane of the sky. Third, any instrument
simulator can be used to transform these pre-observation images into
post-observation images, or the rest-frame spectra of the {\it
Archives} into post-observation observer-frame spectra.  We have
shown in the previous section how {\tt SkyMaker} can be used in this
way, because {\tt FieldMaker} is able to generate the relevant
information in the proper format. Clearly, many different cones can be
generated within a single \gal~{\it Archive} by changing the observing
point, the direction of the line-of-sight and/or the aperture. Many
pre-observing images can be generated from a single cone by changing
the filter response curves. And many post-observing images can be
generated from a single pre-observing image by changing the
instrument simulator. The information produced by a given
\mo~post-processing constitutes what we hereafter call the {\it
Products} of this post-processing.

At this stage, from any single simulation, we have generated a set of
{\it Archives} and {\it Products} that includes tables of halo and
galaxy properties, and FITS files of spectra and images. Two big
issues obviously appear. First, the size of the database makes it very
cumbersome. As an example, for a standard \gal~post-processing of our
$\Lambda$CDM simulation, the total numbers of halos and galaxies
generated in the 70 snapshots respectively amount to 1.5 and 1.8
million. The output {\it Archives} are about 4.5 GB for tables and 45
GB for spectra FITS files, not to speak of the {\it Products}.
Second, the specific information that is relevant for a given user is
hidden within the bulk of non-relevant information.  Let's
imagine for instance that we want to get the $B$-band absolute
magnitude and total cold gas mass of a random sample of 100 galaxies
brighter than apparent magnitude $I_{AB}=20$. Extracting this
information will require reading tables with many columns (the
properties) and many rows (the galaxies). It may be possible to
anticipate the latter issue, and generate many specific tables for
many different situations and potential users, but the same pieces of
information will consequently be duplicated many times, which is not
the proper way to proceed.

The solution to this conundrum is well known: it consists in storing
the information into a {\it relational database} (hereafter
RDB). Here, we use the word database (with its loose meaning) for all
the information we want to make available, and the word relational
database (with its strict meaning) for the technical way of putting
part of this information into a specific structure.

We decided to use MySQL as the relational database server. MySQL is a
freely available, widely used and extremely fast database server which
is capable enough for our purposes. Tools to provide Web-based access
to the MySQL server are also available.  The tables generated by the
\gal~and \mo~post-processing are stored in MySQL {\it tables}. Our database input and testing is done using
several short scripts in Perl that use the Perl/DBI module. The Web
front-end uses PHP4 to pass SQL queries to the MySQL database. Query
outputs can either be displayed as an HTML table within the browser or
down-loaded to a local file.  In this section we briefly describe the
database. A quick-start guide, sample queries and descriptions of the
various fields in each table are available at the \gal~web-site ({\tt http://galics.iap.fr}).

\begin{figure}
\psfig{figure=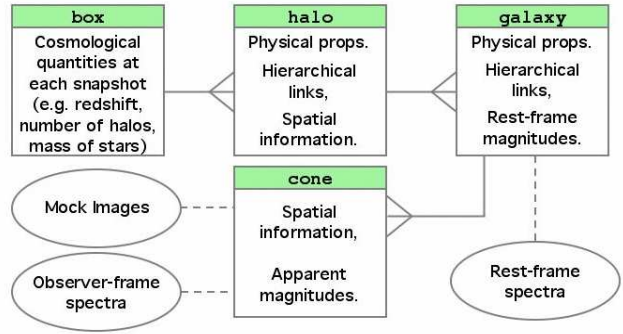,width=\hssize}
\caption{Data model of the \gal/\mo~ database. There are four layers of
information corresponding to four tables. The box table contains
information on mean quantities computed for each time-step, e.g.
redshift, cosmic time, mean SFR. The halo table contains information
on the DM halos of each time-step, including their dynamical properties
and merging history. The galaxy table  contains
information about the physical properties of galaxies at each
time-step, including rest-frame magnitudes and spectra. The cone table
contains information on the mock observing cone, mainly positions and
apparent magnitudes. Companion FITS files include rest-frame spectra,
pre- or post-observing images, and observer-frame spectra.}
\label{fig:database}
\end{figure}

From a single dark matter simulation, each choice in the list of the
input parameters corresponds to a \gal~post-processing with its
specific {\it Archives} and {\it Products}, which in its turn
corresponds to a single MySQL database. The information is stored into
a structure designed after the usual analysis in terms of entities,
attributes and relationships, that is designed to minimise storage
space and maximise query speed.  Each MySQL database is consequently
organised in three MySQL tables for the {\it Archives}, respectively
called the {\it box}, {\it halo}, and {\it galaxy} tables, and
numerous {\it cone} tables for the {\it Products}. The database scheme
is illustrated in Figure \ref{fig:database}.
\begin{enumerate}
\item The box table contains general information about mean quantities
at each snapshot of the simulation, such as the cosmic time,
corresponding redshift, total number of halos and galaxies within the
box, and integrated cosmic quantities such as the cosmic star
formation rate, cold gas content, hot gas content, etc.
\item The halo table contains information on the halos at each
time-step. Each halo is identified by a unique ID in the simulation.
This information deals with dark matter (e.g. mass of the halo, virial
radius, circular velocity) as well as the baryonic content of the
halos (e.g. mass of hot gas and its metalicity). On top of this, we
include spatial information, namely positions and velocities of the
centres of mass of the halos, and hierarchical information, that is,
merging history links. This information is in principle enough for one
to run one's own semi-analytic model on our dark matter simulations,
and thus freely test new recipes and compare results with \gal.
\item The galaxy table contains the physical information we compute
for galaxies: stellar masses, star formation rates, gas contents,
rest-frame absolute magnitudes in a variety of filters, etc. Each
galaxy is identified by a unique ID in the simulation.
\item The cone tables contain the positions of galaxies distributed in
a mock catalogue, along with their apparent magnitudes in a variety of
filters. Each galaxy is identified by a unique ID in the
cone. However, because of transverse replication, different cone
galaxy ID's can point to the same galaxy ID in the simulation. There
are several cone tables, corresponding to different random seeds for
the box shuffling process (which mimics, to some extent, cosmic
variance), or to different field sizes.
\end{enumerate}

Of course, the information included in the four tables is usable
simultaneously in the queries, since the ID's of halos and
galaxies are shared by the tables of the database and allow one
to pass information from one table to another.  Companion information
on mock spectra and images is stored as FITS files.  The rest-frame
spectra are related to the galaxies in the galaxy table, whereas the
images are related to a particular cone table. The observer-frame
spectra are related both to galaxies and to the cone from which the
galaxies are identified.

The \gal{} web-site also contains a {\it hierarchical query} page
which allows the user to retrieve hierarchical information for any
galaxy in mock catalogues or snapshots. At the moment, this page
contains 3 scripts. The first one allows one to view the full merging
tree of a given galaxy, identified by its unique ID. The second one
allow one to follow the evolution of selected properties of a galaxy
along its merging history tree. Here, three options are available :
(i) one can follow properties along the {\it main branch} which links
a galaxy to its most massive progenitor at each merger; (ii) one can
follow the {\it most massive branch}, which links the most massive
progenitors at each timestep together; or (iii) one can perform a sum
of additive properties on all progenitors at each timestep. Each of
these options is a different way to retrieve partial information
contained in the full merging history tree of a galaxy. The third
script allows one to get the list of descendents or progenitors (at
any redshift) of a sample of galaxies selected with any set of
criteria. An example of use of this powerful script can be found in
\citet{BlaizotEtal04}. These scripts allow, for the first time, the
systematic exploration of the evolution of galaxies in the framework
of hierarchical galaxy formation.

\section{Conclusions}
In this paper, we presented the {\it Mock Map Facility} that takes the
results of our \gal~hybrid model of hierarchical galaxy formation to
make mock galaxy samples. Our method involves the construction of
observing cones by integrating through the snapshots of the $N$-body
simulation, and by using the properties of galaxies as they are
computed by the \gal~post-processing. This technique builds up on the
simulation and is affected by the shortcomings of the latter (mass
resolution, and absence of rare objects due to the limited size of the
box). It also incorporates shortcomings due to radial replication
along the line-of-sight, and, for large solid angles, transverse
replication. We introduced box re-shuffling to minimise replication
effects. The price of this technique is the loss of some signal for
the correlation functions (both 2D and 3D) on distances smaller than
the size of the box. This loss is generally not larger than 10\%. Of
course, there is no signal on distances larger than the size of the
box, and finite volume effects have been shown to introduce a
significant (but well understood) bias on angular correlation
functions.

For the purpose of analysing the limitations of our method, we
compared predictions of \gal{}/\mo{} to various observations. We
showed that the model agrees well with $K$-band counts and redshift
distributions. And we showed that within finite volume effects, the
model also agrees well with the APM angular correlation
function. These results have been obtained with the same model as used
in \citet{BlaizotEtal04} which showed good agreement with the
properties of Lyman break galaxies at $z\sim 3$. This shows that our
mock catalogues can readily be used for a variety of scientific
investigations.

From the mock catalogues of the observing cones, we show how to make
``realistic'' mock images. Since our \gal~post-processing involves
multi-wavelength information from the UV to the sub-millimetre range,
our mock images are produced through a wide range of standard filters.
These field images can be observed through any instrument simulator.
The technique is able to produce input lists for {\tt SkyMaker}.
Instrument simulators adapted to observations at infrared and
sub-millimetre wavelengths can also be used. 

The database produced by the \gal~and \mo~post-processing is quite
large, and has to be stored in such a way that easy access to relevant
information is provided. We put the results into a relational database
structure to which SQL queries can be passed through a simple Web
interface. This structure has a number of well-known advantages: it
optimises storage space, it makes access to the relevant information very easy, it is able to deal with simultaneous
queries and updates, etc. The results of \gal~(physical properties,
rest-frame magnitudes) and \mo~(observable properties, apparent
magnitudes) are stored in this database, and linked together through
the standard system of a relational database model. FITS files of mock
images and spectra are also available linked from the database.

The content of the database can be used for several purposes. For
instance:
\begin{itemize}
\item comparison of mock predictions with observations through the
  production of a mock survey that can be processed with the same data
  processing pipeline as the actual survey;
\item elaboration of observing strategies for forthcoming satellite
  missions and ground-based instruments;
\item benchmark for data processing pipelines; A database populated
  with \gal~sources is a valuable ``test set'' on which to base and test
  the various techniques and algorithms for data reduction and analysis,
  for the next generation of astronomical instrumentation; the database
  includes the positions and magnitudes of the galaxies that are put
  into the mock images, and can be used as a ``truth table'' that has to be
  recovered by the data processing software;
\item creation of customised galaxy samples for comparison with other
  models, or observational data.
\end{itemize}

We are considering improvements to this prototype database.  They can
develop along three axes: (i) In the mid-term future, the foreseen
computer performances make replications unavoidable if a sufficient
level of mass resolution for galaxy studies has to be attained.
However, the improvement of the simulations will result in larger
boxes that will decrease the number of radial and transverse
replications, and be able to include rarer objects.  (ii) The
improvement of the physics within the simulations will also make
better mass resolution possible, and will (hopefully) produce better
results.  There is no doubt also that the semi-analytic recipes have
to be improved. The same cone building technique will be used also for
converting the outputs of $N$-body simulations $+$ hydrodynamics into
mock observations. (iii) The database prototype that has been
presented here will be enhanced to make it compatible with the data
and metadata standards that are now being developed as part of the
theoretical virtual observatory. The present \mo~will form a valuable
test-bed for testing the integration of theoretical data from
simulations into the theoretical virtual observatory, which forms a
part of the global Astronomical Virtual Observatory effort.

\section*{acknowledgements}
The authors thank the anonymous referee for insightful comments that
helped improve this paper. YW was supported by projects 1610-1 and
1910-1 of the Indo-French Centre For Advanced Scientific Research
(CEFIPRA). The N-body simulation used in this work was run on the Cray
T3E at the IDRIS supercomputing facility.


\begin{thebibliography}{}

\bibitem[\protect\citeauthoryear{{Benson}, {Cole}, {Frenk}, {Baugh} \&
  {Lacey}}{{Benson} et~al.}{2000}]{BensonEtal00}
{Benson} A.~J.,  {Cole} S.,  {Frenk} C.~S.,  {Baugh} C.~M.,    {Lacey} C.~G.,
  2000, \mnras, 311, 793

\bibitem[\protect\citeauthoryear{{Berlind} \& {Weinberg}}{{Berlind} \&
  {Weinberg}}{2002}]{BerlindWeinberg02}
{Berlind} A.~A.,  {Weinberg} D.~H.,  2002, \apj, 575, 587

\bibitem[\protect\citeauthoryear{{Bernardeau}, {Colombi}, {Gaztanaga} \&
  {Scoccimarro}}{{Bernardeau} et~al.}{2002}]{BernardeauEtal02}
{Bernardeau} F.,  {Colombi} S.,  {Gaztanaga} E.,    {Scoccimarro} R.,  2002,
  \physrep, 367, 1

\bibitem[\protect\citeauthoryear{{Bertin} \& {Arnouts}}{{Bertin} \&
  {Arnouts}}{1996}]{BertinArnouts96}
{Bertin} E.,  {Arnouts} S.,  1996, \aaps, 117, 393

\bibitem[\protect\citeauthoryear{{Blaizot}, {Guiderdoni}, {Devriendt},
  {Bouchet}, {Hatton} \& {Stoehr}}{{Blaizot} et~al.}{2004}]{BlaizotEtal04}
{Blaizot} J.,  {Guiderdoni} B.,  {Devriendt} J.~E.~G.,  {Bouchet} F.~R.,
  {Hatton} S.~J.,    {Stoehr} F.,  2004, \mnras, 352, 571

\bibitem[\protect\citeauthoryear{{Bouwens}, {Broadhurst} \& {Silk}}{{Bouwens}
  et~al.}{1998}]{BouwensBroadhurstSilk98}
{Bouwens} R.,  {Broadhurst} T.,    {Silk} J.,  1998, \apj, 506, 557

\bibitem[\protect\citeauthoryear{{Cimatti}, {Pozzetti}, {Mignoli}, {Daddi},
  {Menci}, {Poli}, {Fontana}, {Renzini}, {Zamorani}, {Broadhurst}, {Cristiani},
  {D'Odorico}, {Giallongo} \& {Gilmozzi}}{{Cimatti}
  et~al.}{2002}]{CimattiEtal02}
{Cimatti} A.,  {Pozzetti} L.,  {Mignoli} M.,  {Daddi} E.,  {Menci} N.,  {Poli}
  F.,  {Fontana} A.,  {Renzini} A.,  {Zamorani} G.,  {Broadhurst} T.,
  {Cristiani} S.,  {D'Odorico} S.,  {Giallongo} E.,    {Gilmozzi} R.,  2002,
  \aap, 391, L1

\bibitem[\protect\citeauthoryear{Cimatti et
al.}{2002}]{CimattiEtal02a} Cimatti A., et al., 2002, A\&A, 392, 395

\bibitem[\protect\citeauthoryear{{Coil}, {Davis} \& {Szapudi}}{{Coil}
  et~al.}{2001}]{CoilDavisSzapudi01}
{Coil} A.~L.,  {Davis} M.,    {Szapudi} I.,  2001, \pasp, 113, 1312

\bibitem[\protect\citeauthoryear{Eke, Cole, \&
Frenk}{1996}]{EkeColeFrenk96} Eke V.~R., Cole S., Frenk C.~S., 1996,
MNRAS, 282, 263 

\bibitem[\protect\citeauthoryear{{Cole}, {Hatton}, {Weinberg} \&
  {Frenk}}{{Cole} et~al.}{1998}]{ColeEtal98}
{Cole} S.,  {Hatton} S.,  {Weinberg} D.~H.,    {Frenk} C.~S.,  1998, \mnras,
  300, 945

\bibitem[\protect\citeauthoryear{{Colless} et~al.,}{{Colless}
  et~al.}{2001}]{CollessEtal01}
{Colless} M.,  et~al., 2001, \mnras, 328, 1039

\bibitem[\protect\citeauthoryear{{Davis}, {Efstathiou}, {Frenk} \&
  {White}}{{Davis} et~al.}{1985}]{DavisEtal85}
{Davis} M.,  {Efstathiou} G.,  {Frenk} C.~S.,    {White} S.~D.~M.,  1985, \apj,
  292, 371

\bibitem[\protect\citeauthoryear{{Devriendt}, {Guiderdoni} \&
  {Sadat}}{{Devriendt} et~al.}{1999}]{DevriendtGuiderdoniSadat99}
{Devriendt} J.~E.~G.,  {Guiderdoni} B.,    {Sadat} R.,  1999, \aap, 350, 381

\bibitem[\protect\citeauthoryear{{Diaferio}, {Kauffmann}, {Colberg} \&
  {White}}{{Diaferio} et~al.}{1999}]{DiaferioEtal99}
{Diaferio} A.,  {Kauffmann} G.,  {Colberg} J.~M.,    {White} S.~D.~M.,  1999,
  \mnras, 307, 537

\bibitem[\protect\citeauthoryear{{Erben}, {Van Waerbeke}, {Bertin}, {Mellier}
  \& {Schneider}}{{Erben} et~al.}{2001}]{ErbenEtal01}
{Erben} T.,  {Van Waerbeke} L.,  {Bertin} E.,  {Mellier} Y.,    {Schneider} P.,
   2001, \aap, 366, 717

\bibitem[\protect\citeauthoryear{{Evrard}, {MacFarland}, {Couchman}, {Colberg},
  {Yoshida}, {White}, {Jenkins}, {Frenk}, {Pearce}, {Peacock} \&
  {Thomas}}{{Evrard} et~al.}{2002}]{EvrardEtal02}
{Evrard} A.~E.,  {MacFarland} T.~J.,  {Couchman} H.~M.~P.,  {Colberg} J.~M.,
  {Yoshida} N.,  {White} S.~D.~M.,  {Jenkins} A.,  {Frenk} C.~S.,  {Pearce}
  F.~R.,  {Peacock} J.~A.,    {Thomas} P.~A.,  2002, \apj, 573, 7

\bibitem[\protect\citeauthoryear{{G{\' o}rski}, {Banday}, {Hivon} \&
  {Wandelt}}{{G{\' o}rski} et~al.}{2002}]{GorskiEtal02}
{G{\' o}rski} K.~M.,  {Banday} A.~J.,  {Hivon} E.,    {Wandelt} B.~D.,  2002,
  in ASP Conf. Ser. 281: Astronomical Data Analysis Software and Systems XI
  {HEALPix --- a Framework for High Resolution, Fast Analysis on the Sphere}.
pp 107--+

\bibitem[\protect\citeauthoryear{{Hatton}, {Devriendt}, {Ninin}, {Bouchet},
  {Guiderdoni} \& {Vibert}}{{Hatton} et~al.}{2003}]{HattonEtal03}
{Hatton} S.,  {Devriendt} J.~E.~G.,  {Ninin} S.,  {Bouchet} F.~R.,
  {Guiderdoni} B.,    {Vibert} D.,  2003, \mnras, 343, 75

\bibitem[\protect\citeauthoryear{{Helly}, {Cole}, {Frenk}, {Baugh}, {Benson} \&
  {Lacey}}{{Helly} et~al.}{2003}]{HellyEtal03a}
{Helly} J.~C.,  {Cole} S.,  {Frenk} C.~S.,  {Baugh} C.~M.,  {Benson} A.,
  {Lacey} C.,  2003, \mnras, 338, 903

\bibitem[\protect\citeauthoryear{{Hernquist}}{{Hernquist}}{1990}]{Hernquist90}
{Hernquist} L.,  1990, \apj, 356, 359

\bibitem[\protect\citeauthoryear{{Kauffmann}, {Colberg}, {Diaferio} \&
  {White}}{{Kauffmann} et~al.}{1999}]{KauffmannEtal99a}
{Kauffmann} G.,  {Colberg} J.~M.,  {Diaferio} A.,    {White} S.~D.~M.,  1999,
  \mnras, 303, 188

\bibitem[\protect\citeauthoryear{{Kauffmann}, {Nusser} \&
  {Steinmetz}}{{Kauffmann} et~al.}{1997}]{KauffmannNusserSteinmetz97}
{Kauffmann} G.,  {Nusser} A.,    {Steinmetz} M.,  1997, \mnras, 286, 795

\bibitem[\protect\citeauthoryear{{Kennicutt}}{{Kennicutt}}{1983}]{Kennicutt83}
{Kennicutt} R.~C.,  1983, \apj, 272, 54

\bibitem[\protect\citeauthoryear{{Landy} \& {Szalay}}{{Landy} \&
  {Szalay}}{1993}]{LandySzalay93}
{Landy} S.~D.,  {Szalay} A.~S.,  1993, \apj, 412, 64

\bibitem[\protect\citeauthoryear{{Madau}}{{Madau}}{1995}]{Madau95}
{Madau} P.,  1995, \apj, 441, 18

\bibitem[\protect\citeauthoryear{{Maddox}, {Efstathiou} \&
  {Sutherland}}{{Maddox} et~al.}{1996}]{MaddoxEfstathiouSutherland96}
{Maddox} S.~J.,  {Efstathiou} G.,    {Sutherland} W.~J.,  1996, \mnras, 283,
  1227

\bibitem[\protect\citeauthoryear{{Ninin}}{{Ninin}}{1999}]{Ninin99}
{Ninin} S.,  1999, PhD thesis, Universit\'e Paris 11

\bibitem[\protect\citeauthoryear{Norberg et 
al.}{2002}]{NorbergEtal02} Norberg P., et al., 2002, MNRAS, 336, 907 

\bibitem[\protect\citeauthoryear{{Peacock} \& {Smith}}{{Peacock} \&
  {Smith}}{2000}]{PeacockSmith00}
{Peacock} J.~A.,  {Smith} R.~E.,  2000, \mnras, 318, 1144

\bibitem[\protect\citeauthoryear{{Scoccimarro}, {Sheth}, {Hui} \&
  {Jain}}{{Scoccimarro} et~al.}{2001}]{ScoccimarroEtal01}
{Scoccimarro} R.,  {Sheth} R.~K.,  {Hui} L.,    {Jain} B.,  2001, \apj, 546, 20

\bibitem[\protect\citeauthoryear{{Somerville}, {Lemson}, {Sigad}, {Dekel},
  {Kauffmann} \& {White}}{{Somerville} et~al.}{2001}]{SomervilleEtal01}
{Somerville} R.~S.,  {Lemson} G.,  {Sigad} Y.,  {Dekel} A.,  {Kauffmann} G.,
  {White} S.~D.~M.,  2001, \mnras, 320, 289

\bibitem[\protect\citeauthoryear{{Steidel} \& {Hamilton}}{{Steidel} \&
  {Hamilton}}{1993}]{SteidelHamilton93}
{Steidel} C.~C.,  {Hamilton} D.,  1993, \aj, 105, 2017

\bibitem[\protect\citeauthoryear{{Yang}, {Mo}, {Jing}, {van den Bosch} \&
  {Chu}}{{Yang} et~al.}{2004}]{YangEtal04}
{Yang} X.,  {Mo} H.~J.,  {Jing} Y.~P.,  {van den Bosch} F.~C.,    {Chu} Y.,
  2004, \mnras, 350, 1153

\end{thebibliography}

\appendix

\section{Using the database}
In this appendix, we briefly illustrate how the database can be used to
interpret observational data in the paradigm of hierarchical galaxy
formation. We give four examples which exemplify the kind of
information that can be exploited:

\begin{enumerate}
\item synthesis of a {\it volume--limited sample} of galaxies, for
instance at $z\simeq 0$,
\item synthesis of a {\it magnitude--limited sample} of galaxies, and
related multi-wavelength information,
\item search for {\it 2D and 3D spatial information} (e.g for redshift
distribution, clustering), and correlation of properties with it,
\item search for {\it merging history trees} within hierarchical
galaxy formation (e.g. in what type of galaxy does the material of a
given high--redshift galaxy end up at $z\simeq 0$?  How many
progenitors does a galaxy at $z\simeq 0$ have?).
\end{enumerate}

For each example, we give a typical SQL query that returns the
requested subsample by querying the database. We refer the reader to the Web
page ({\tt http://galics.iap.fr/}) for additional examples, and a simple introduction to SQL syntax.

\subsection{Volume--limited samples}
It is possible to query the database to list a
series of physical properties for a subsample of galaxies with
sophisticated selection criteria. As an example, select 100
galaxies at random in the $z=0$ snapshot (that corresponds to timestep
70), with the requirement that their Johnson $B$--band absolute
magnitude is brighter than $-20$, and their dispersion velocity is
larger than 200 km/s. We are also interested in obtaining their absolute
$K$--band magnitude, $B-K$ colour, morphological types and total stellar mass.

\noindent
{\tt  > SELECT gal\_id, type\_B2D\_lum, tot\_JOHNSON\_B,\\
      > tot\_speed,  tot\_JOHNSON\_K,\\ 
      > tot\_JOHNSON\_B-tot\_JOHNSON\_K, tot\_mstar\\
      > FROM galaxy\\
      > WHERE timestep=70\\
      > AND tot\_JOHNSON\_B < -20 \\
      > AND tot\_speed > 200\\
      > ORDER BY RAND()\\
      > LIMIT 100}\\
The last two commands
place the list of galaxies recovered by the query in random order, and then limit the output to the first 100 rows. This example takes less than 1 second to run.

\subsection{Magnitude--limited samples}
Another type of query  is to select galaxies according to
their apparent magnitudes in order to mimic an observational sample
or to predict what a  forthcoming survey will yield. The selected
sample can then be studied in the explicit cosmological context of
\gal, and the physical properties of the selected galaxies can be
retrieved easily to gain insight on the nature of the ``observed''
objects. An example of using mock catalogues to interpret
observational data is given in \citet{BlaizotEtal04}.  A crucial issue
in observational galaxy formation studies is the identification of
counterparts at any wavelength of galaxies observed through any given
filter. It is often quite a challenge, for example, to identify the
optical counterparts of far infrared sources, observed with low
angular resolution. The \gal~database provides a powerful tool to
address these questions as it predicts emission properties of galaxies
from the UV to the sub-mm and gives the opportunity to build
corresponding mock maps.

An example SQL query to retrieve a magnitude limited sample in a 1
deg$^2$ cone is given below, for galaxies brighter than $I_{AB} =
22.5$. We are interested in their apparent B-K colour (in the AB
system), their total stellar mass, and the virial mass of their host
halo. Such a query requires information that is present not only in
the cone table, but also in the galaxy table and the halo table.  It
requires what is called a {\it join} in SQL syntax:

\noindent
{\tt  > SELECT cone\_001.cone\_id, cone\_001.app\_redshift, \\
  > cone\_001.JOHNSON\_BAB, cone\_001.JOHNSON\_KAB, \\
  > cone\_001.JOHNSON\_BAB-cone\_001.JOHNSON\_KAB,  \\
  > galaxy.tot\_mstar, halo.m\_vir \\
  > FROM galaxy, cone\_001, halo \\
  > WHERE cone\_001.JOHNSON\_IAB < 22.5 \\
  > AND cone\_001.gal\_id = galaxy.gal\_id\\
  > AND halo.halo\_id=galaxy.halo\_id}\\

This example query runs in about 30 seconds, and returns information (7 columns) for about 31000 galaxies.

The same type of selection can be used to work the other way around: one
can select galaxies according to their physical properties or their
dark matter halo properties, and extract their spatial distribution
and apparent magnitudes from the cone.

\subsection{Spatial information}
Spatial information can be retrieved the same way as above, for galaxies in mock catalogues. Consider the following query~: 

\noindent
{\tt > SELECT right\_ascension, declination, app\_redshift\\ 
     > FROM cone\_001 \\
     > WHERE JOHNSON\_IAB < 22.5}\\

This produces a table with the angular coordinates and apparent
redshifts of all the mock galaxies brighter than 22.5 in the I$_{AB}$
band within a 1 deg$^2$ field. There are again about 31000 such galaxies, and the query here runs in about 5 seconds. 

\subsection{Hierarchical evolution}
The \gal~project gives for the first time the opportunity to interpret
observational data within the paradigm of hierarchical galaxy
formation in a systematic way. One of the most important features of
this theoretical framework is the notion of galaxy merging history
tree.  Going up or down this tree allows one to investigate the
properties of the progenitors or descendents of any given galaxy at
any redshift, as well as the mass build--up of that galaxy.  In the
galaxy table, each galaxy has a pointer towards its unique descendant
at the next timestep.  This minimal information is sufficient to
reconstruct any merging history tree, whether forward (the list of
descendants, that is a single branch as time flows) or backwards (the
list of progenitors, that may be a full tree with many branches as we
look back). The number of merging events for the progenitors of the
galaxy whose ID is (the character string) xxyyyyyzzz is easily
obtained through the following query, as well as the ID of its
descendant at the next timestep (for all timesteps but the last
one):

\noindent
{\tt > SELECT nb\_merge, daughter\_num FROM galaxy WHERE gal\_id='xxyyyyyzzz'}\\

It is necessary to run the above query recursively to build up the full merging history trees. On the \gal~website we provide PHP scripts that generate such recursive queries and pass them to the database server. 
Once a
galaxy ID is supplied by the user through the Web interface, the ID's of its
progenitors and descendants, as well as their properties, are
recovered through the ``recursive query'' page.  An interesting option
allows the user to obtain the sum of the (additive) properties
of all the progenitors. In such a way, the evolution of the total Star
Formation Rate or the total stellar mass in all the progenitors can be
easily followed. Examples of such recursive queries can be found in
\devriendta~and \blaizota.

\end{document}